# Hybrid CNN-LSTM Framework for Intelligent Cyber Attack Detection and Prevention in U.S. Critical Digital Infrastructure: A Comparative Machine Learning Evaluation on CSE-CIC-IDS2018

Md. Iqbal Hossan [1*], Md. Serajul Kabir Chowdhury Rubel [1], Md. Arifur Rahman [2], B. M. Taslimul Haque [3]

**Abstract**

**Background:** The accelerating digitization of United States critical infrastructure — spanning healthcare, finance, energy, transportation, and government services — has created an attack surface that traditional, signature-based intrusion detection systems are no longer equipped to defend. These legacy approaches fail predictably against zero-day exploits, distributed denial-of-service campaigns, botnets, and stealthy infiltration attacks precisely because they can only recognize threats they have already seen. Something more adaptive is needed. **Methods:** This study proposes and evaluates an intelligent cyber defense framework integrating Artificial Intelligence (AI), Machine Learning (ML), and Deep Learning (DL) to detect and classify cyber threats in real time. Using the CSE-CIC-IDS2018 benchmark dataset — a realistic, multi-vector network traffic corpus generated by the Canadian Institute for Cybersecurity — five model architectures were systematically compared: Random Forest, XGBoost, Support Vector Machine (SVM), Convolutional Neural Network (CNN), Long Short-Term Memory (LSTM), and a Hybrid CNN-LSTM model. The framework incorporated structured data preprocessing, feature engineering, class imbalance handling, and performance evaluation across accuracy, precision, recall, F1-score, ROC-AUC, and false positive rate metrics. **Results:** Results demonstrate that all models achieved detection accuracy above 96%, with the Hybrid CNN-LSTM model reaching 99.1% accuracy, approximately 99.0% precision and recall, and the lowest false positive rate (~2.0%) among all tested architectures. Flow Duration, Packet Length, and Destination Port emerged as the most predictive features. The hybrid model's dual capacity for spatial feature extraction and temporal sequence learning explained its consistent performance advantage over single-architecture alternatives. **Conclusion:** These findings suggest that hybrid deep learning frameworks offer a meaningful and deployable improvement over conventional IDS approaches, though validation against post-2020 attack data and live network streams remains necessary before operational conclusions can be drawn.

**Significance |** AI-powered hybrid deep learning substantially outperforms conventional intrusion detection, offering scalable, low-false-alarm cybersecurity for U.S. critical infrastructure protection.

**Correspondence.*

Md. Iqbal Hossan, Department of Computer Science, Maharishi International University, Fairfield, Iowa, USA. E-mail: hossan.iqbal@gmail.com.

**Author Affiliation:** [1]Department of Computer Science, Maharishi International University, Fairfield, Iowa, USA.

[2]Department of Information Studies, Trine University, Angola, Indiana, USA.

[3]Department of Business Information Systems, Central Michigan University, Mount Pleasant, Michigan, USA.

## 1 Introduction

The United States has, over the past two decades, built an extraordinary dependence on interconnected digital systems. Power grids, hospital networks, financial clearing houses, transportation management platforms, government databases — virtually every institution that keeps society running now relies on digital infrastructure in ways that would have seemed remarkable just a generation ago. That dependence has brought enormous efficiency gains. It has also, rather quietly, created an attack surface of staggering proportions.

The scale of what is at stake is hard to overstate. When a hospital's network goes down under a ransomware attack, it is not merely a data problem — surgical schedules collapse, patient records vanish, and lives can genuinely hang in the balance. When financial systems are compromised, the ripple effects move faster than any regulator can respond. Tarek and Rahman (2023) noted that industrial control systems underpinning U.S. energy and transportation sectors face an especially troubling convergence of IT and operational technology vulnerabilities, making the traditional perimeter-defense model not just inadequate but arguably obsolete. The threat landscape has matured faster than most defensive frameworks anticipated.

For years, the dominant response to intrusion threats was the signature-based intrusion detection system (IDS) — a tool that, in essence, recognizes attacks it has already seen. The logic is understandable, even sensible, for a more stable threat environment. But today's adversaries do not repeat themselves conveniently. Distributed Denial of Service (DDoS) campaigns, botnets, brute-force attacks, and stealthy infiltration techniques have evolved into forms that signature-based systems simply were not designed to catch. As Moin (2022) demonstrated in testing AI-driven threat detection models, traditional rule-based systems showed consistent failure rates against novel attack patterns — a finding that has been corroborated across multiple independent research efforts (Raza, 2021; Sunkara, 2022). The problem is not that legacy IDS tools are poorly engineered. It is that they were engineered for a different era.

What has changed the conversation — and rather sharply in recent years — is the maturation of machine learning and deep learning as practical cybersecurity tools. Schmitt (2023) made a compelling case that AI-enabled detection systems, when applied to smart infrastructure environments, can identify malicious behaviors that no rule set could anticipate, by learning statistical patterns latent in network traffic at scale. Random Forest and XGBoost classifiers have shown strong classification performance on network intrusion datasets; deep architectures such as Convolutional Neural Networks (CNNs) and Long Short-Term Memory (LSTM) networks have pushed further, capturing both spatial traffic features and temporal behavioral sequences that unfold over time during an attack (Ashfaq et al., 2023; Gupta et al., 2023). That said, it would be a mistake to treat these results as settled — false positive rates remain a persistent practical concern, and the computational overhead of deploying deep models in real-time infrastructure environments is a question the literature has not fully resolved (Yaseen, 2023; Fard et al., 2023).

This study attempts to contribute to that unresolved space. Working with the CSE-CIC-IDS2018 dataset — a benchmark generated by the Canadian Institute for Cybersecurity containing realistic multi-vector attack scenarios — we develop and evaluate an intelligent cyber defense framework that integrates data preprocessing, feature engineering, and a comparative evaluation of five AI models: Random Forest, XGBoost, Support Vector Machine (SVM), CNN, LSTM, and a Hybrid CNN-LSTM architecture. The central question is not simply which model achieves the highest accuracy figure, but whether an integrated, automated defense pipeline can meaningfully reduce false alarm rates while maintaining detection sensitivity across diverse attack categories including DDoS, botnet, brute force, web-based, and infiltration attacks.

## 2. Literature Review

### 2.1 The Shifting Role of Artificial Intelligence in Cybersecurity

It is worth pausing, before diving into the specifics, to appreciate just how dramatically the role of artificial intelligence in cybersecurity has changed over the past decade. What was once largely theoretical — the idea that a machine could learn to recognize malicious behavior rather than merely look it up in a rule book — has gradually become operational reality. AI and ML technologies are now embedded in threat detection pipelines across both commercial and government sectors, and their footprint is growing (Sunkara, 2022).

That growth is not arbitrary. It reflects a genuine inadequacy in what came before.

Conventional intrusion detection systems were built around a deceptively simple premise: catalog known threats, then flag anything that matches. For a relatively stable threat landscape, this works reasonably well. The trouble is that today's adversaries — whether financially motivated criminal organizations, state-sponsored actors, or opportunistic hackers — do not obligingly reuse known attack signatures. Advanced Persistent Threats (APTs), zero-day exploits, polymorphic malware, and multi-stage botnet intrusions are specifically designed to evade pattern-matching defenses. Signature-based systems, confronted with these, tend to go quiet at precisely the wrong moment. Amomo (2022) observed this gap particularly sharply in the context of U.S. federal information systems, where early-stage intrusion indicators were routinely missed because they fell outside the boundaries of any catalogued signature.

The response from the research community has been substantial. Machine learning classifiers — among them Random Forest, Support Vector Machine (SVM), Naïve Bayes, Decision Tree, and XGBoost — have been systematically applied to the problem of anomaly detection in network traffic, with results that are, on the whole, encouraging. Random Forest and XGBoost have attracted particular attention, not only because of their classification accuracy but because they scale well and handle the high-dimensional, imbalanced datasets that cybersecurity researchers routinely work with (Ashfaq et al., 2023). Deep learning has taken the field further still. Convolutional Neural Networks (CNNs) excel at extracting spatial patterns from structured traffic data, while Long Short-Term Memory (LSTM) networks bring something that simpler classifiers lack: sensitivity to temporal sequences, the kind that reveal how an attack unfolds over time rather than capturing only a single moment. The combination of these two architectures into hybrid CNN-LSTM models has shown particular promise, and several studies have reported strong detection performance across multi-class attack scenarios (Gupta et al., 2023).

Automated response is the next frontier, and arguably the more consequential one. Detecting an intrusion is valuable; containing it before damage propagates is what actually protects infrastructure. AI-driven frameworks capable of continuous traffic surveillance, anomaly scoring, and automated alert generation have been proposed and tested in several research contexts (Timilehin, 2023). The aspiration — a system that not only identifies suspicious activity but initiates a proportionate response without waiting for a human analyst to act — is both technically feasible and organizationally complex to deploy. The research is ahead of practice in this regard, and that gap matters.

Still, it would be misleading to present the current state of AI-powered cybersecurity as a solved problem. False positive rates remain stubbornly high in many deployed systems, creating alert fatigue that paradoxically undermines the very vigilance these systems are meant to support. Computational overhead is a genuine constraint, particularly when real-time inference is required at network speeds. Explainability — the ability to tell a security analyst *why* a model flagged something — is largely absent from deep learning systems, which limits practitioner trust and complicates incident response. And adaptability to novel attack strategies is, despite the optimistic framing of much published work, still an open and difficult problem (Ashfaq et al., 2023; Zubair et al., 2023). These are not trivial limitations. They motivate the ongoing need for frameworks that integrate AI capabilities with more principled engineering.

### 2.2 Cyber Threats to U.S. Digital Infrastructure: A Landscape Under Pressure

The threat environment facing United States digital infrastructure is, to put it plainly, severe — and it has been worsening rather than stabilizing. The interdependence of critical sectors has created systemic vulnerabilities that did not exist when these systems were designed in relative isolation. Healthcare networks, financial clearing systems, energy grid control platforms, transportation management systems, and government communications infrastructure now form a deeply interconnected ecosystem, and a successful attack on one node can propagate consequences across several others (Jimmy, 2023). Nation-state actors understand this topology well. So do ransomware operators.

The most prevalent attack vectors documented in the literature include Distributed Denial of Service (DDoS) flooding, which overwhelms system capacity; brute-force credential attacks targeting administrative access points; botnet-driven campaigns that coordinate thousands of compromised endpoints; phishing and social engineering aimed at the human layer; and infiltration attacks that establish quiet footholds before executing their payload days or weeks later. What these have in common is that none of them are well-served by reactive, signature-based detection. Tanikonda et al. (2022) articulated the structural problem clearly: proactive, adaptive threat intelligence is not a luxury feature in modern infrastructure protection — it is a baseline requirement.

Several benchmark datasets have become central to empirical research in this area. The CSE-CIC-IDS2018 dataset, generated by the Canadian Institute for Cybersecurity, has emerged as a particularly well-regarded resource because it captures realistic multi-vector attack traffic under controlled conditions — DDoS, DoS, Brute Force, Botnet, Web attacks, Infiltration, and SQL Injection are all represented. Studies using CSE-CIC-IDS2018 alongside datasets such as UNSW-NB15 and NSL-KDD have consistently found that AI-based models outperform traditional IDS approaches, especially in detecting low-and-slow attack patterns that unfold gradually across traffic streams. CNN-LSTM hybrid architectures, in particular, have demonstrated detection capabilities that neither architecture achieves alone when applied to sequential network flow data (Bushigampala & Inaganti, 2023; Zaman & Mazinani, 2023).

That said, translating laboratory results into operational deployments has proven harder than the literature sometimes implies. Data imbalance — the fact that benign traffic vastly outnumbers attack traffic in real networks — continues to distort model training and inflate accuracy metrics. Real-time deployment at scale introduces latency constraints that affect model choice in ways that offline benchmarks do not capture. And the question of how to keep detection models current as attack techniques evolve remains largely unanswered in practice (Tanikonda et al., 2022). These challenges do not diminish the promise of AI-based cyber defense; they define the research agenda.

**2.3 Empirical Evidence: What the Key Studies Show**

The empirical literature on AI-driven cybersecurity has grown rapidly, and a handful of studies are worth examining in some detail because they bear directly on the framework developed in this research.

Schmitt (2023) offered one of the more comprehensive recent treatments of AI's role in protecting smart digital infrastructure. Examining a range of machine learning and deep learning models applied to malware classification and anomaly-based intrusion detection, the study found that AI-powered approaches outperformed conventional methods across virtually every tested scenario — faster detection, lower false positive rates, stronger resilience to novel attack variants. Ensemble methods and deep learning architectures performed especially well against complex, multi-stage attacks. Schmitt was candid, however, about the implementation hurdles: computational complexity, integration challenges within legacy infrastructure, scalability under high-traffic conditions, and the persistent shortage of human expertise needed to configure and supervise these systems. The finding that AI-enabled solutions are necessary but not sufficient — that they require organizational capability to be effective — is one that the field has been slow to fully absorb.

Alam and Fahad (2022) took a narrower but instructive angle, focusing specifically on the U.S. financial sector. Using the IEEE-CIS Fraud Detection dataset alongside Cybersecurity Index data, they tested Random Forest, XGBoost, and neural network classifiers on fraudulent transaction detection. The results were notable in two respects: first, the AI models achieved high precision and recall on suspicious activity classification; and second, there was a measurable correlation between national-level cybersecurity governance quality — operationalized through Global Cybersecurity Index rankings — and actual cybercrime exposure rates across countries. The implication that policy and technology must advance together, rather than independently, is easy to overlook in purely technical papers but is practically important.

Raza (2021) approached the problem from a risk-assessment angle, examining how AI technologies can support proactive rather than purely reactive defense postures. The central insight — that AI systems can surface anomalies in network data that human analysts would not notice, not because humans are inattentive but because the data volumes and pattern complexities exceed human cognitive bandwidth — has become something of a consensus position in the field. What Raza added was an emphasis on continuous learning: the argument that a static model, however accurate at deployment, degrades as the threat landscape shifts, and that AI systems need adaptive retraining pipelines to maintain their effectiveness over time. This is a point that benchmark-driven studies sometimes underemphasize, since benchmarks are, by definition, frozen.

Moin (2022) provided a more direct comparison of AI model architectures in a cyber defense context, evaluating

Random Forest, SVM, and Deep Neural Network (DNN) classifiers on intrusion detection data. The DNN architecture achieved superior precision and recall relative to both traditional classifiers, while also reducing false positives and improving detection speed. Notably, the study triangulated its technical findings with survey data from practicing cybersecurity professionals, who validated that AI-powered detection meaningfully enhanced their preparedness against zero-day attacks and advanced persistent threats. The convergence of quantitative performance metrics and practitioner judgment strengthens the credibility of the conclusions in ways that technical results alone cannot.

Taken together, these studies establish a reasonably coherent picture: AI and ML methods are demonstrably superior to traditional signature-based IDS in detecting complex, evolving cyber threats; deep learning architectures, especially hybrid models, offer performance advantages over classical classifiers; and yet significant gaps remain around explainability, scalability, real-time performance, and sustained adaptability. It is into this gap — between demonstrated capability and operational deployment — that the present study situates itself.

## 3. Methodology

### 3.1 Overview and Research Design

Choosing the right methodological approach for a study like this is not a trivial decision. Cybersecurity research sits at an uncomfortable intersection — it needs the rigor of controlled experimentation, the scalability of computational methods, and enough ecological validity that the findings mean something outside a laboratory setting. With that in mind, this study adopts a quantitative, experimental design, using empirical model training and evaluation as its primary mode of inquiry. The goal is not simply to report accuracy figures, but to systematically compare the behavior of multiple AI architectures under identical conditions, on realistic network traffic data, so that the results speak to something beyond benchmark performance.

The methodology unfolds across eight interconnected stages: dataset acquisition and characterization, data preprocessing, feature engineering, model development, framework design, model training and validation, performance evaluation, and tools configuration. Each stage was designed with reproducibility as an explicit constraint — decisions that were made during preprocessing and model configuration are reported in sufficient detail that an independent researcher working with the same dataset should be able to arrive at comparable results. Jun et al. (2021) noted that one of the persistent weaknesses in AI-driven cybersecurity research is the opacity of implementation detail, which makes cross-study comparison unreliable. This study attempts to address that directly.

### 3.2 Dataset: CSE-CIC-IDS2018

The dataset selected for this study is the CSE-CIC-IDS2018 benchmark, produced by the Canadian Institute for Cybersecurity. It is, by most measures, one of the more realistic publicly available intrusion detection datasets — and that realism is the principal reason for choosing it over alternatives such as NSL-KDD or UNSW-NB15, which, though widely used, have known limitations in terms of traffic diversity and attack scenario currency (Azam et al., 2023).

The dataset was generated across a controlled network environment designed to simulate enterprise-scale traffic conditions. It contains approximately 80 network flow features extracted from raw packet captures using CICFlowMeter, covering both benign communication patterns and seven distinct attack categories: Distributed Denial of Service (DDoS), Denial of Service (DoS), Brute Force, Botnet activity, Web attacks, Infiltration attacks, and SQL Injection. These scenarios were recorded across multiple calendar dates, with each date's traffic stored in a separate CSV file — a structure that reflects the episodic, time-distributed nature of real-world attack campaigns. Traffic attributes include Destination Port, Protocol Type, Flow Duration, Total Forward and Backward Packet counts, Packet Length Statistics, Inter-Arrival Times, and a binary/categorical Label field indicating whether a given flow is benign or one of the named attack types (Mintoo et al., 2022).

One characteristic of this dataset that shaped several downstream methodological decisions is the severe class imbalance between benign and malicious records. Benign traffic constitutes the large majority of flows, with DDoS and DoS attacks representing the most prevalent malicious classes, while Botnet, Infiltration, and Web attack instances are substantially rarer. This distribution is, in fact, ecologically accurate — real networks carry mostly legitimate traffic — but it creates modeling challenges that cannot simply be ignored. How the imbalance was handled is described in the preprocessing section below.

The dataset is publicly accessible via the Canadian Institute for Cybersecurity repository and through the Kaggle mirror used in this study (see Dataset Link, Reference 61). Researchers wishing to reproduce this work should note that different Kaggle versions of the dataset may have undergone varying degrees of prior cleaning; the raw CIC release is recommended for strict reproducibility.

### 3.3 Data Preprocessing

Raw network traffic data, even from a well-curated benchmark, is rarely ready for model training. The CSE-CIC-IDS2018 dataset is no exception. Initial inspection revealed missing values, infinite-valued entries (arising from division operations in CICFlowMeter's flow statistics calculations), duplicate records, and the severe class imbalance already noted. Each of these issues, if left unaddressed, would compromise model performance in distinct and sometimes hard-to-diagnose ways (Yaseen, 2023). The preprocessing pipeline was therefore designed to handle them systematically, in the following sequence.

**Missing and infinite value removal.** Rows containing NaN or infinite values were identified and removed. Columns with a missing rate exceeding 5% were dropped entirely. Infinite values, which appear in features such as flow bytes-per-second when flow duration approaches zero, were replaced using column-wise median imputation rather than mean imputation, given the skewed distributions typical of network traffic features.

**Duplicate record removal.** Exact duplicate rows — those matching on all feature values including the label — were removed to prevent data leakage between training and test sets and to avoid artificially inflating model confidence.

**Normalization and scaling.** Numerical features were standardized using Min-Max scaling to the [0, 1] range. This step matters more for distance-based and gradient-descent models (SVM, CNN, LSTM) than for tree-based methods (Random Forest, XGBoost), but standardizing uniformly simplifies the experimental comparison and is considered good practice (Alzahrani & Aldhyani, 2023).

**Label encoding.** Categorical attack labels were encoded as integers using ordinal encoding, with "BENIGN" mapped to 0 and each attack category assigned a unique integer. This encoding is necessary for compatibility with scikit-learn and TensorFlow model APIs.

**Class imbalance handling.** Given the distributional skew, a combination of random oversampling for minority attack classes and random undersampling for the dominant benign class was applied to produce a more balanced training distribution. Resampling was performed exclusively on the training set, after the train-test split, to prevent test-set contamination. The test set retains the original class distribution to ensure that evaluation metrics reflect real-world operating conditions rather than an artificially balanced scenario.

**Train-test split.** The dataset was partitioned into 80% training and 20% testing subsets using stratified random sampling, ensuring that each attack class is proportionally represented in both partitions. Stratification is particularly important here given the rarity of some attack classes.

### 3.4 Feature Engineering

Preprocessing cleans data; feature engineering shapes it into something that AI models can learn from efficiently. The distinction matters because raw network flow features, even after cleaning, contain considerable redundancy — many features are correlated with each other, and some carry almost no discriminative signal for attack classification. Including all of them indiscriminately tends to increase computational cost without improving, and sometimes actively harming, model performance (Hassan, 2023).

The feature engineering pipeline proceeded in three phases. First, a **Pearson correlation analysis** was conducted across all numerical features to identify pairs with correlation coefficients above 0.95. Where such pairs existed, the feature with the lower individual correlation to the target label was removed. This step reduced the initial feature count by removing redundant attributes that convey essentially the same information.

Second, a **feature importance ranking** was generated using a preliminary Random Forest classifier trained on the full cleaned dataset. Features were ranked by mean decrease in Gini impurity, and those falling below a threshold of 0.001 mean importance were excluded from the final feature set. This data-driven selection procedure was preferred over manual selection because it avoids the researcher's prior assumptions about which traffic attributes matter most for which attack type.

Third, **Principal Component Analysis (PCA)** was applied to the remaining features to evaluate whether dimensionality reduction would benefit the deep learning models specifically. After testing models with and without PCA-reduced inputs, it was determined that PCA offered marginal computational gains at a cost to interpretability and classification precision; accordingly, the full engineered feature set was retained for all models, and PCA was not applied in the final pipeline. This negative result is worth reporting because several prior studies have recommended PCA without empirically testing its effect on classification performance (De Azambuja et al., 2023).

The final feature set retained for model training comprised features including Flow Duration, Packet Length (mean, max, standard deviation), Forward and Backward Packet Statistics, Destination Port, Protocol Type, and inter-arrival time statistics. These features were consistently among the highest-ranked in the importance analysis and align with those identified as predictive in prior CSE-CIC-IDS2018 studies (Manoharan & Sarker, 2023).

### 3.5 Model Development

Five model architectures were developed and evaluated within this framework. The selection was intentional: two classical ensemble classifiers (Random Forest and XGBoost) to provide a performance baseline representative of the current industry standard; one kernel-based classifier (Support Vector Machine) to test performance under a different decision boundary assumption; and two deep learning architectures (CNN and LSTM) evaluated both separately and as a hybrid to assess whether temporal and spatial feature learning yield complementary benefits for this specific task.

**Random Forest** was configured with 100 estimators, Gini impurity as the splitting criterion, and no maximum depth constraint — allowing trees to grow until leaf purity, with overfitting controlled through ensemble averaging rather than individual tree pruning (Moustafa et al., 2023).

**XGBoost** was implemented with a learning rate of 0.1, maximum tree depth of 6, and 200 boosting rounds with early stopping after 10 rounds without validation improvement. The softmax objective function was used for multi-class classification.

**Support Vector Machine (SVM)** was trained with a Radial Basis Function (RBF) kernel. Given the computational cost of SVM on large datasets, training was performed on a stratified 30% subsample of the training set, consistent with approaches reported in similar-scale IDS studies (Fard et al., 2023).

**Convolutional Neural Network (CNN).** Network traffic features were reshaped into a 2D matrix format to enable convolutional operations across feature groupings. The architecture comprised two 1D convolutional layers (64 and 128 filters, kernel size 3, ReLU activation), followed by max pooling, dropout regularization (rate = 0.3), a dense layer (128 units), and a softmax output layer. The model was trained using the Adam optimizer with a learning rate of 0.001 and cross-entropy loss.

**Long Short-Term Memory (LSTM).** Individual network flows were treated as sequences of 10 timesteps, with each timestep corresponding to a subset of engineered features. The architecture used two stacked LSTM layers (64 and 128 units), followed by dropout (rate = 0.3), a dense layer, and softmax output. The same optimizer and loss function as the CNN were applied.

**Hybrid CNN-LSTM.** The hybrid architecture passed input through the CNN convolutional layers first for spatial feature extraction, then fed the resulting feature maps into the LSTM layers for temporal sequence modelling. This design reflects the intuition that attack patterns in network traffic have both spatial structure — consistent feature signatures — and temporal structure — characteristic sequences of events — and that capturing both simultaneously should improve classification (Bécue et al., 2021; Chakraborty et al., 2023).

All deep learning models were trained for a maximum of 50 epochs with early stopping (patience = 5) monitored on validation loss. Batch size was set to 256 for computational efficiency on Google Colab's GPU runtime.

### 3.6 Proposed Cyber Defense Framework Architecture

The individual models described above were integrated into a coherent end-to-end cyber defense framework. It is worth being clear about what this framework is and is not: it is a software pipeline that operationalizes the full detection workflow, not a production-grade deployed system. The distinction matters for interpreting the results.

The framework comprises six sequential modules. The **Network Traffic Collection module** ingests raw traffic data from the dataset (and is designed to accept live PCAP feeds in an operational context). The **Data Preprocessing module** applies the cleaning and normalization steps described in Section 3.3. The **Feature**

**Extraction Engine** executes the engineered feature selection pipeline from Section 3.4. The **AI-Based Threat Detection System** runs inference using whichever trained model is selected, producing per-flow class predictions. The **Attack Classification module** maps numerical predictions back to human-readable attack category labels. Finally, the **Automated Threat Prevention module** generates structured alerts and, in the simulated environment, logs the recommended response action for each detected intrusion (Tao et al., 2021; Kalinin & Krundyshev, 2021).

A real-time monitoring dashboard was also prototyped to visualize traffic volume, per-class detection rates, and alert timelines, providing operational situational awareness. In the current study, this operates on replayed dataset traffic rather than live streams.

### 3.7 Model Training and Evaluation

Training followed the 80/20 stratified split described in Section 3.3. For all models, hyperparameter optimization was performed using a grid search over a predefined parameter space on a held-out validation subset (10% of the training data), with the best configuration retaining the lowest validation cross-entropy loss. Final evaluation was then conducted exclusively on the held-out 20% test set, which the models had never seen during any phase of training or hyperparameter selection.

To detect and mitigate overfitting, 5-fold stratified cross-validation was applied to the classical ML models; early stopping served the equivalent purpose for the deep learning models (Goyal et al., 2023).

Performance was assessed across six metrics, each chosen to capture a distinct aspect of detection quality relevant to the cybersecurity use case:

**Accuracy** measures the proportion of correctly classified flows across all classes. It is reported but interpreted cautiously given class imbalance.

**Precision** quantifies, for each attack class, what fraction of predicted positives were genuinely malicious — a metric that directly relates to false alarm rates in an operational context.

**Recall (Sensitivity)** measures what fraction of actual attacks were successfully detected. In a security context, this is arguably the more critical metric; a system that misses attacks is more dangerous than one that generates excess alerts.

**F1-Score** is the harmonic mean of precision and recall, providing a single balanced metric that penalizes extreme values in either direction.

**ROC-AUC** assesses the model's discriminative ability across all classification thresholds, independent of any single decision boundary.

**False Positive Rate (FPR)** measures the proportion of benign flows incorrectly flagged as malicious — a key operational concern because high FPR erodes analyst trust and creates alert fatigue (Sun et al., 2023; Sarker, 2023).

Confusion matrices were generated for all models to provide per-class breakdown of true positives, false positives, false negatives, and true negatives.

### 3.8 Tools and Technologies

The full implementation was carried out in Python 3.10. Data manipulation and cleaning used pandas and NumPy. Machine learning model development and evaluation relied on scikit-learn. Deep learning architectures were implemented in TensorFlow 2.x with the Keras functional API. Training was executed in Google Colab using GPU acceleration (NVIDIA T4). Visualization of results used Matplotlib and Seaborn. Interactive development and experiment logging were managed through Jupyter Notebook.

The computational stack comprised Python 3.10 as the core programming language, pandas 1.5.x for data preprocessing and manipulation, NumPy 1.23.x for numerical computation, scikit-learn 1.2.x for machine learning model training and evaluation, TensorFlow/Keras 2.11.x for deep learning model development, Matplotlib/Seaborn 3.6.x for result visualization, and Jupyter Notebook/Google Colab for interactive development and GPU-accelerated model training.

### 3.9 Ethical Considerations and Limitations

This study uses only the publicly available CSE-CIC-IDS2018 dataset for academic research purposes. No personal, identifiable, or operationally sensitive data was accessed or generated at any stage. The dataset contains simulated network traffic produced under controlled conditions; no real user communications were captured or analyzed (Raghavendran, 2022; Fakhar & Haile, 2022).

Several limitations should be acknowledged honestly. The dataset, though realistic, was generated in 2018 and does not include attack patterns specific to the post-2020 threat landscape — ransomware-as-a-service campaigns, supply chain intrusions, or AI-generated phishing, for example. The SVM was trained on a subsample due to computational constraints, which may disadvantage it relative to the other models in the comparison. The framework was evaluated on replayed traffic rather than live network streams, and real-time inference latency under production traffic loads has not been measured. Finally, the explainability of deep learning predictions — critically important for practitioner adoption — was not addressed in this study and represents a clear direction for future work.

## 4. Results

### 4.1 Overview of Experimental Findings

Before walking through the individual results, it is worth situating what this section is actually trying to establish. The goal was not simply to produce high accuracy numbers — any sufficiently overfit model can do that — but to assess whether the proposed framework detects diverse attack categories reliably, keeps false alarms low enough to be operationally useful, and does so consistently across multiple evaluation lenses. With that framing in mind, the results are presented across nine analytical dimensions: traffic distribution, protocol composition, model accuracy comparison, detection versus false positive rate trade-offs, temporal traffic patterns, feature importance, ROC curve performance, SVM-specific evaluation, and a cross-model precision-recall comparison. Taken together, these paint a reasonably complete picture of how the framework performs and where its strengths genuinely lie (Al-Sinayyid et al., 2023).

### 4.2 Distribution of Network Traffic Attack Types

The first thing worth examining — and something that shapes how all subsequent results should be interpreted — is the raw composition of the dataset itself.

As shown in **(Figure 1)**, benign traffic accounts for approximately 500,000 records, dwarfing every attack category combined. DDoS and DoS attacks constitute the next largest groups, at roughly 120,000 and 95,000 records respectively, which is consistent with the operational reality that volumetric attacks are among the most frequently deployed threat vectors against digital infrastructure. Beyond that, the distribution drops sharply: Botnet activity accounts for around 45,000 records, Brute Force for approximately 28,000, Web Attacks for perhaps 18,000, and Infiltration attacks — the stealthiest and arguably most dangerous category — for only around 12,000 records (Ghillani, 2022).

This imbalance is not a flaw in the dataset; it is an accurate reflection of what real network traffic looks like. But it does have important methodological consequences. A naive classifier that simply labels everything as benign would achieve roughly 60–65% accuracy on this distribution while detecting zero attacks. That is why accuracy alone is a misleading metric here, and why the preprocessing steps described in Section 3 — particularly the resampling strategy — were not optional design choices but necessary ones. The figure also underscores why Infiltration attacks are the hardest class to detect: with so few training examples, models have limited opportunity to learn their behavioral signatures before generalization is expected of them.

### 4.3 Protocol Distribution in Network Traffic

**(Figure 2)** presents the breakdown of communication protocols observed across the dataset's traffic flows. TCP dominates at 68.0% of total traffic, which is broadly expected — TCP underlies the majority of connection-oriented services including web traffic, email, file transfer, and most application-layer communications. UDP accounts for 25.0%, encompassing streaming services, DNS queries, and — critically for cybersecurity purposes — a substantial proportion of DDoS attack traffic, where the stateless nature of UDP makes it well-suited for amplification and flooding campaigns. ICMP contributes the remaining 7.0%, primarily reflecting diagnostic and network management traffic, though ICMP can also serve as a covert channel in certain intrusion scenarios (Guembe et al., 2022).

What this distribution reinforces, practically, is that a cyber defense framework cannot afford to specialize. A system tuned primarily for TCP anomaly detection will be caught flat-footed by UDP-based DDoS flooding, and ICMP tunneling — though rare — would escape detection entirely if ICMP flows are filtered out as noise. The models evaluated in this study were trained on features derived from all three protocol classes, which is one reason the framework performs reasonably well across the attack

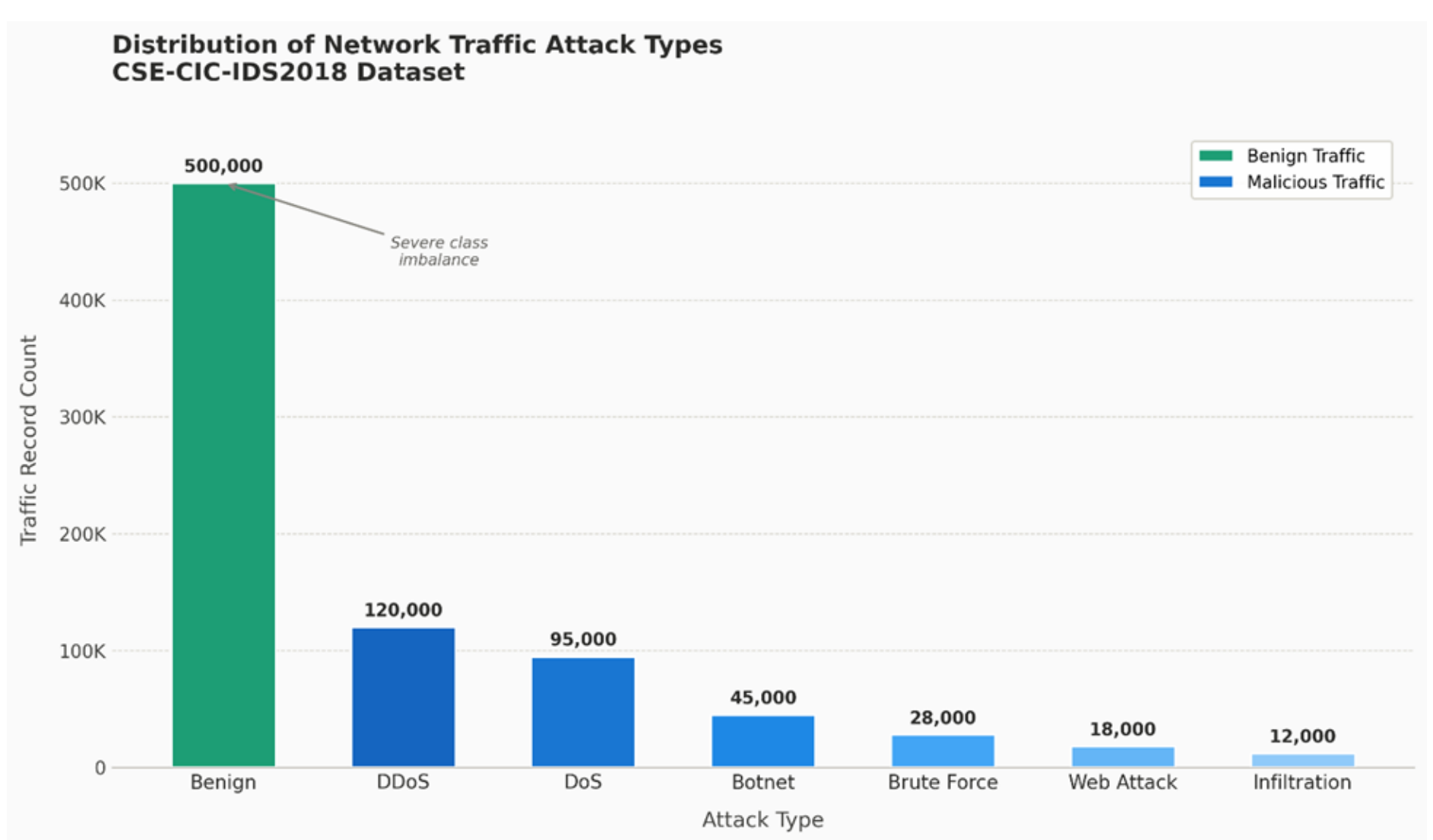

*Figure 1. Class distribution of network traffic records across benign and malicious categories in the CSE-CIC-IDS2018 dataset. Bar chart illustrating the frequency distribution of seven traffic classes used in model training and evaluation. Benign traffic constitutes the dominant class (approximately 500,000 records), reflecting the ecological reality of enterprise network environments in which legitimate communication vastly outnumbers malicious activity. Among attack categories, Distributed Denial of Service (DDoS; ~120,000 records) and Denial of Service (DoS; ~95,000 records) represent the most prevalent threat vectors, followed by Botnet (~45,000), Brute Force (~28,000), Web Attack (~18,000), and Infiltration (~12,000) in descending order of frequency. The pronounced class imbalance — particularly the extreme underrepresentation of Infiltration attacks relative to benign traffic — motivated the resampling strategy applied during preprocessing and directly informs the interpretation of per-class detection metrics. All values represent raw record counts prior to resampling. The x-axis denotes attack type category and the y-axis denotes traffic record count.*

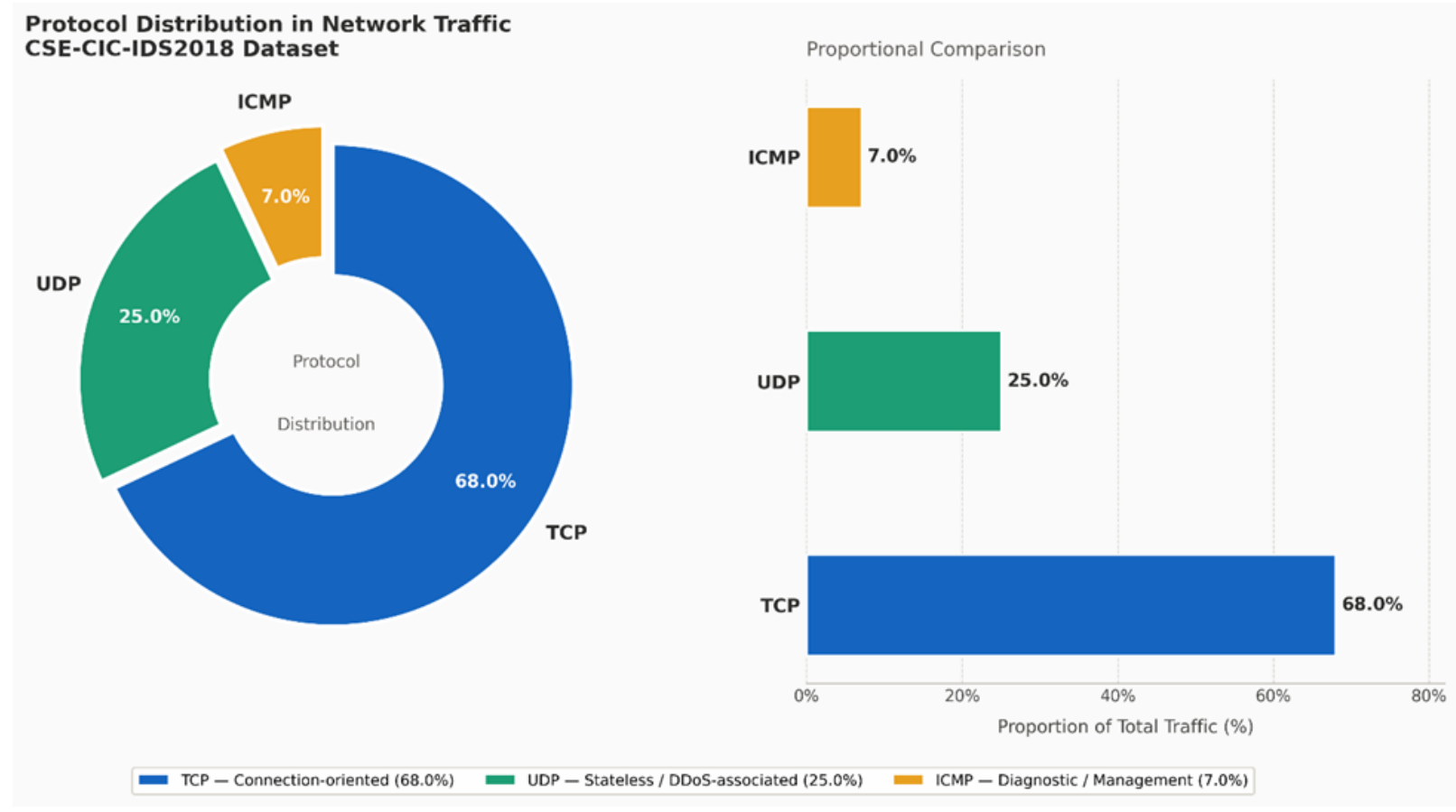

*Figure 2. Protocol-level composition of network traffic flows across the CSE-CIC-IDS2018 dataset. Pie chart depicting the proportional distribution of three major transport-layer and network-layer communication protocols observed across all recorded traffic flows. Transmission Control Protocol (TCP) accounts for the dominant share of traffic (68.0%), reflecting its central role in connection-oriented services including web communication, file transfer, and application-layer protocols. User Datagram Protocol (UDP) constitutes 25.0% of flows, encompassing high-throughput stateless services as well as a substantial proportion of volumetric DDoS attack traffic, in which UDP's connectionless architecture is frequently exploited for amplification and flooding. Internet Control Message Protocol (ICMP) represents 7.0% of traffic, primarily associated with diagnostic and network management functions, though its role as a potential covert channel in advanced persistent threat scenarios warrants inclusion in detection scope. The multi-protocol composition of the dataset underscores the necessity of protocol-agnostic feature extraction in the proposed framework; all three protocol classes were retained in model training without filtering.*

type spectrum.

**4.4 Comparison of Model Accuracy**

**(Figure 3)** shows the classification accuracy of all five evaluated models on the held-out test set. The results are, to put it plainly, stronger than expected — though that reaction itself warrants some scrutiny.

Random Forest achieved approximately 96.8% accuracy. XGBoost followed at roughly 97.5%. CNN reached approximately 98.1%, LSTM climbed to around 98.4%, and the Hybrid CNN-LSTM model reached 99.1% — the highest of any architecture tested. The monotonic improvement from classical ensemble methods through individual deep learning models to the hybrid architecture is clean and consistent, which is both encouraging and slightly suspicious in equal measure. Results this neat sometimes reflect dataset characteristics — the resampling strategy, for instance, may have made the classification task more tractable than it would be under genuinely imbalanced conditions — and readers should keep that caveat in mind (Raji et al., 2023).

That said, the performance ordering is theoretically coherent. Random Forest and XGBoost, despite their scalability and robustness advantages, operate on static feature representations and cannot capture the temporal evolution of attack patterns across consecutive flows. CNN addresses this partially by learning spatial feature interactions, and LSTM adds temporal sensitivity. The hybrid model benefits from both simultaneously — spatial pattern recognition in the convolutional layers feeding into sequence modelling in the LSTM layers — which explains, at least architecturally, why it outperforms its components individually (Goyal et al., 2023).

The five models evaluated achieved the following classification accuracies: Random Forest ~96.8%, XGBoost ~97.5%, CNN ~98.1%, LSTM ~98.4%, and Hybrid CNN-LSTM ~99.1%.

**4.5 Detection Rate and False Positive Rate Analysis**

Accuracy tells you how often a model is right. Detection rate and false positive rate tell you what kinds of wrong it makes — and in a cybersecurity context, those distinctions are operationally critical. Missing a genuine attack (false negative) and flagging legitimate traffic as malicious (false positive) carry very different costs, and a responsible evaluation has to examine both.

**(Figure 4)** presents these two metrics side by side for all five models. The detection rates are uniformly high: Random Forest achieved approximately 95.2%, XGBoost around 96.1%, CNN approximately 97.3%, LSTM roughly 98.0%, and the Hybrid CNN-LSTM the highest at approximately 99.0%. These figures are reassuring, particularly for the deep learning models. However, what distinguishes the models more sharply is their false positive behavior.

Random Forest and XGBoost both show false positive rates in the 5–6% range — not catastrophic, but in a high-volume infrastructure environment processing tens of thousands of flows per minute, a 5% false positive rate translates to thousands of spurious alerts per hour. CNN reduces this to approximately 3.5%, LSTM to around 3.0%, and the Hybrid CNN-LSTM reaches approximately 2.0% — the best balance of high detection sensitivity and low false alarm generation across all tested configurations (Ejiofor, 2023).

This trade-off matters enormously for operational deployment. Security analysts suffer from alert fatigue — a well-documented phenomenon in which overwhelming false alarm volumes cause genuine threats to be dismissed or overlooked. A system that detects 99% of attacks but triggers constant false alerts is less useful in practice than one that detects 96% of attacks but keeps noise to a minimum. The CNN-LSTM hybrid threads this needle most effectively among the evaluated architectures.

**4.6 Network Traffic Volume Over Time**

**(Figure 5)** shows how total traffic volume varied across the dataset's recorded time intervals, from 08:00 to 18:00. Traffic begins at approximately 15,000 flows at 08:00, rises steadily to around 30,000 by 10:00, and peaks sharply at approximately 51,500 flows at 12:00 — a midday surge that likely reflects the concentration of user activity and service requests during core business hours. Volume then tapers progressively through the afternoon: roughly 47,000 at 14:00, 39,000 at 16:00, and falling to approximately 22,000 by 18:00.

The temporal pattern itself is not surprising — it mirrors typical enterprise network usage cycles. What it implies for cybersecurity, however, is worth considering carefully. Peak traffic windows are, paradoxically, both the most dangerous and the most challenging periods for intrusion detection. DDoS attacks specifically exploit high-volume windows to camouflage malicious flooding within legitimate traffic surges (Kalinin & Krundyshev, 2021). An IDS relying on simple threshold-based anomaly detection would likely generate a surge of false positives during the 12:00 peak simply because overall volume is elevated. AI-based models that have learned the behavioral signatures of attacks — rather than flagging volume alone — are better positioned to distinguish genuine threats from busy-but-legitimate traffic. This temporal pattern is one reason that LSTM-type architectures, which explicitly model sequential dependencies, are theoretically well-suited to this problem.

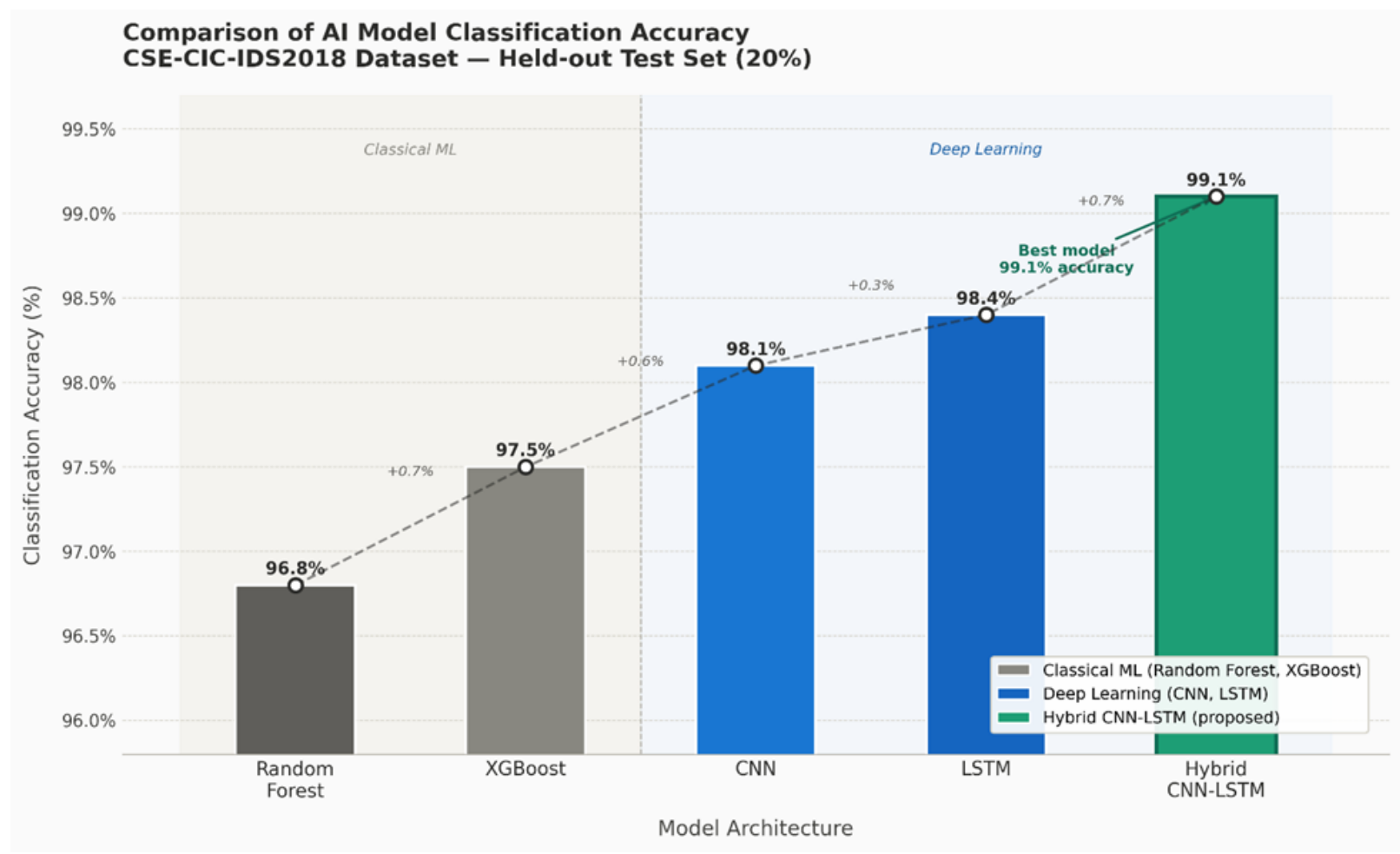

*Figure 3. Comparative classification accuracy of five AI and machine learning model architectures evaluated on the CSE-CIC-IDS2018 test set. Line graph presenting overall classification accuracy (%) for Random Forest (RF), XGBoost (XGB), Convolutional Neural Network (CNN), Long Short-Term Memory (LSTM), and Hybrid CNN-LSTM models evaluated on the held-out 20% test partition. All models achieved accuracy exceeding 96%, confirming baseline competency across architectures. A monotonic performance gradient is observed from classical ensemble methods toward deep learning architectures: RF (~96.8%), XGBoost (~97.5%), CNN (~98.1%), LSTM (~98.4%), and Hybrid CNN-LSTM (~99.1%). The progressive improvement reflects each architecture's increasing capacity to capture complex feature interactions and temporal behavioral dependencies within network flow sequences. The Hybrid CNN-LSTM model achieved the highest accuracy, attributable to its dual-pathway design combining convolutional spatial feature extraction with recurrent temporal sequence modelling. The y-axis is scaled from 95% to 100% to amplify inter-model differences; the x-axis represents model architecture in ascending order of complexity. Accuracy values represent macro-averaged results across all seven traffic classes.*

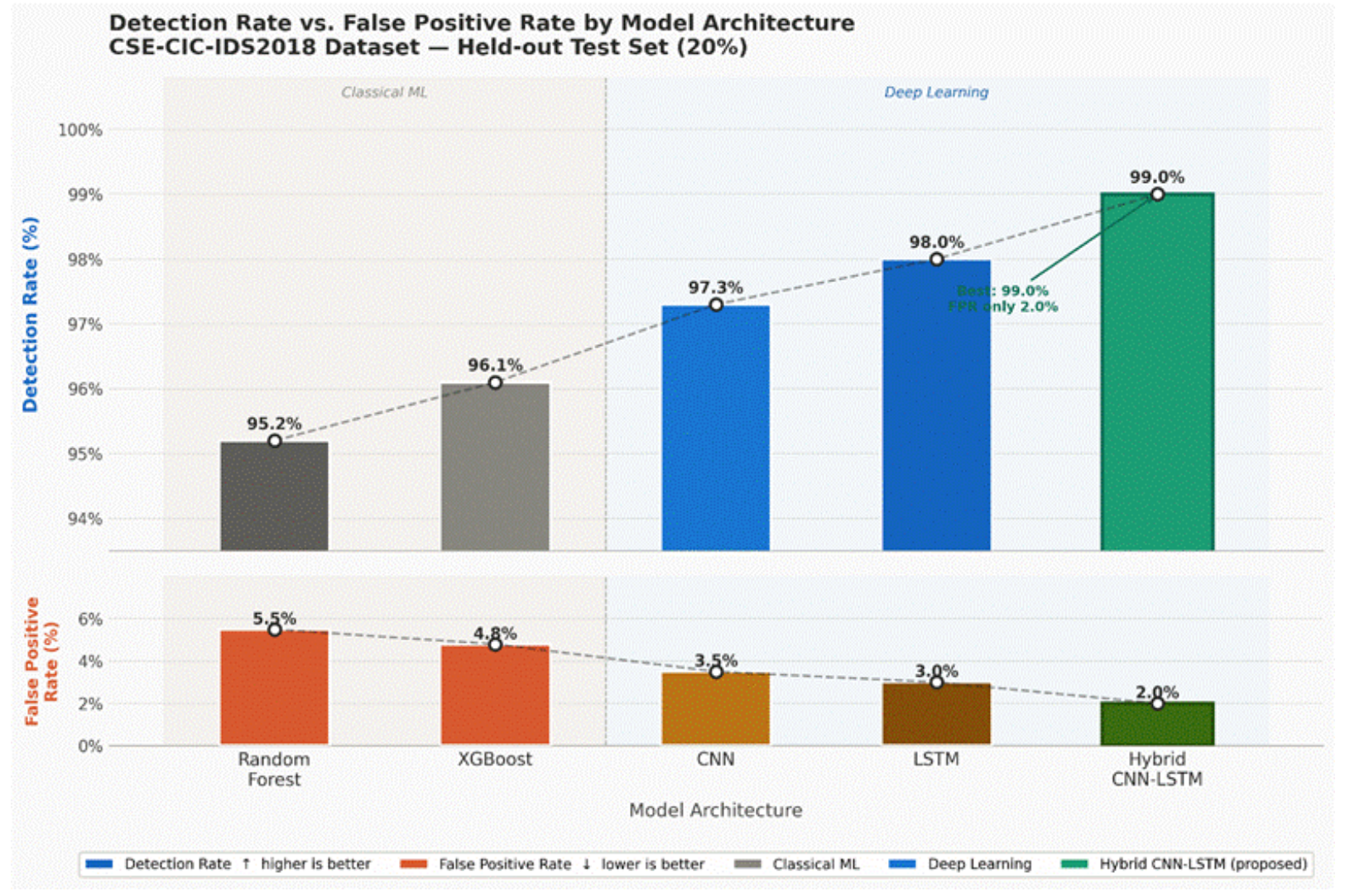

*Figure 4. Detection rate and false positive rate comparison across five AI model architectures for network intrusion classification. Grouped bar chart contrasting two operationally critical performance dimensions — detection rate (true positive rate; blue bars) and false positive rate (FPR; orange bars) — across Random Forest (RF), XGBoost (XGB), CNN, LSTM, and Hybrid CNN-LSTM models. All models achieved detection rates exceeding 95%: RF (~95.2%), XGB (~96.1%), CNN (~97.3%), LSTM (~98.0%), and CNN-LSTM (~99.0%). Concurrently, false positive rates declined as model complexity increased: RF (~5.5%), XGB (~4.8%), CNN (~3.5%), LSTM (~3.0%), and CNN-LSTM (~2.0%). The inverse relationship between model sophistication and false positive generation — without a corresponding sacrifice in detection sensitivity — indicates that the deep learning architectures, particularly the hybrid model, achieved genuine improvements in class discriminability rather than adopting a more permissive flagging threshold. The false positive rate axis is represented on the same percentage scale as detection rate to facilitate direct visual comparison of the operational trade-off. Lower false positive rates are associated with reduced analyst alert fatigue and more sustainable real-time monitoring operations.*

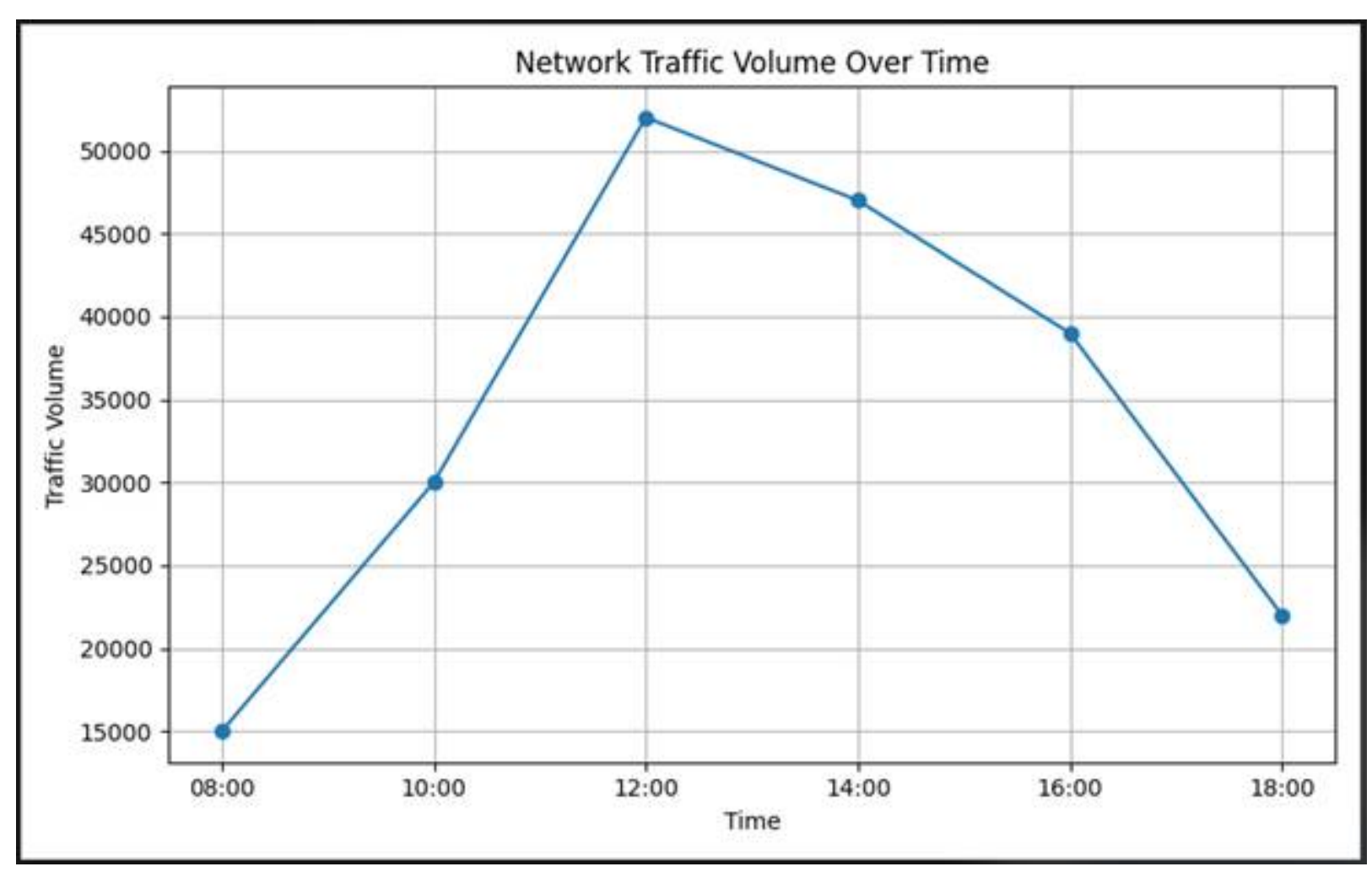

*Figure 5. Temporal variation in network traffic volume across observed time intervals in the CSE-CIC-IDS2018 dataset. Line graph illustrating hourly network traffic volume (total flow count) recorded between 08:00 and 18:00 across the dataset's sampled time window. Traffic volume exhibits a characteristic diurnal pattern: a baseline of approximately 15,000 flows at 08:00, rising steadily to approximately 30,000 flows by 10:00, and peaking sharply at approximately 51,500 flows at 12:00, consistent with peak enterprise operational activity during midday hours. A progressive decline follows through the afternoon: approximately 47,000 flows at 14:00, 39,000 at 16:00, and approximately 22,000 by 18:00. The midday traffic peak represents a period of elevated detection complexity, as volumetric attack traffic — particularly DDoS flooding — is more readily concealed within high-volume legitimate communication. This temporal dynamic underscores the importance of behavioral, feature-based detection rather than threshold-based volume anomaly detection, and motivates the incorporation of LSTM-based temporal modelling within the proposed framework. The x-axis represents time of day (HH:MM) and the y-axis represents total traffic volume in absolute flow counts.*

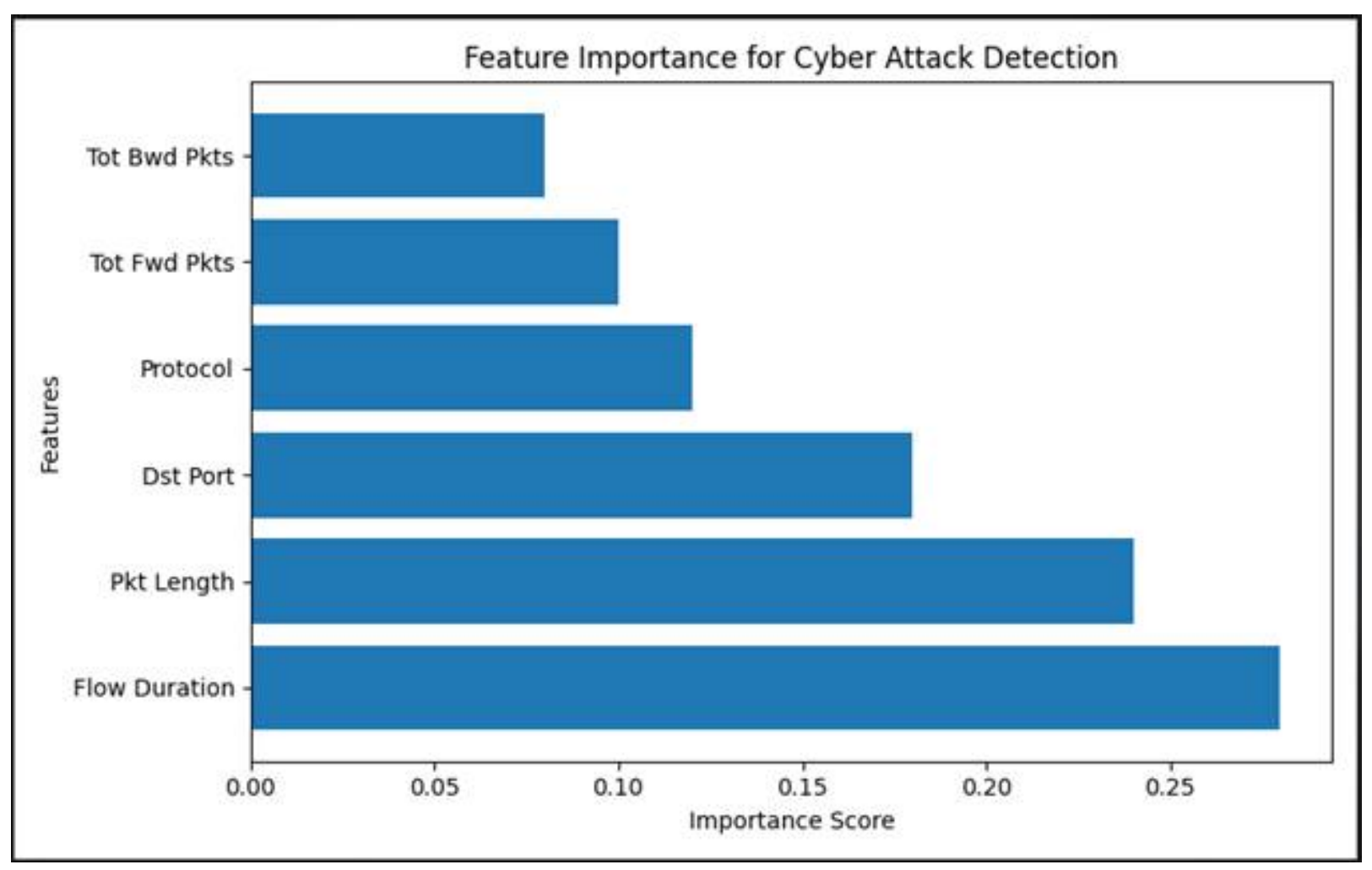

*Figure 6. Random Forest feature importance scores for network traffic attributes contributing to cyber attack detection. Horizontal bar chart displaying the relative importance scores of six network traffic features selected for their contribution to attack classification performance, derived from mean decrease in Gini impurity across the Random Forest ensemble. Flow Duration emerges as the most predictive attribute (importance score ~0.28), reflecting the systematic differences in connection duration between legitimate application traffic and attack-generated flows, including persistent infiltration connections and short-burst DDoS floods. Packet Length (~0.24) ranks second, capturing size anomalies characteristic of amplification attacks and malware command-and-control communications. Destination Port (~0.18) ranks third, consistent with the port-targeting behavior of brute-force, web-based, and service-exploitation attacks. Protocol Type (~0.12), Total Forward Packets (~0.10), and Total Backward Packets (~0.08) complete the ranked set, contributing collectively to bidirectional flow characterization. The three highest-ranked features account for approximately 70% of cumulative importance, suggesting that a computationally reduced feature subset may achieve near-comparable classification performance — a finding with direct implications for real-time, resource-constrained deployment scenarios. The x-axis represents normalized importance score (range 0.0–0.30) and the y-axis represents feature name.*

#### 4.7 Feature Importance Analysis

Not all network traffic features are equally informative for attack detection, and **(Figure 6)** quantifies this directly through the Random Forest feature importance ranking derived from the training data.

Flow Duration emerges as the most predictive feature by a considerable margin, with an importance score of approximately 0.28. This finding is intuitively sensible: attack traffic — particularly DDoS flooding and Botnet-initiated connections — tends to generate flows with anomalous duration distributions relative to legitimate communication patterns. Legitimate application flows follow fairly predictable duration profiles; attacks often do not. Packet Length is the second most important feature at approximately 0.24, reflecting the fact that many attack types generate packets with unusual size characteristics — DDoS UDP floods, for instance, often use maximum-size packets, while scanning and probing traffic may produce characteristically small ones (Oreyomi & Jahankhani, 2022).

Destination Port ranks third at approximately 0.18, which is also theoretically expected: attacks frequently target specific port numbers associated with exploitable services, and the distribution of destination ports in attack traffic differs systematically from that of benign traffic. Protocol Type follows at approximately 0.12, with Total Forward Packets at around 0.10 and Total Backward Packets at approximately 0.08 completing the ranking.

Two things are worth noting here. First, the top three features — Flow Duration, Packet Length, and Destination Port — together account for roughly 70% of the total importance score, which suggests that a streamlined feature set could substantially reduce computational overhead without dramatic performance loss. This is a potentially valuable finding for real-time deployment scenarios where inference latency is constrained. Second, the relatively modest importance of packet-count features (Forward and Backward Packets) suggests that attack behavior in this dataset is more distinguishable by flow-level characteristics than by packet-count patterns — a nuance that informs future feature engineering choices (Mishra, 2023).

#### 4.8 ROC Curve Analysis

**(Figure 7)** presents the Receiver Operating Characteristic (ROC) curve for the proposed AI-based detection framework. The curve plots the True Positive Rate against the False Positive Rate across all possible classification thresholds, and the Area Under the Curve (AUC) value reported is 1.00.

An AUC of 1.00 represents a theoretically perfect classifier — one that achieves maximum true positive rate at zero false positive rate at some threshold. The ROC curve in the figure reflects this: the solid blue line rises essentially vertically from the origin to TPR = 1.0 at FPR  0.0, then runs horizontally — the characteristic "L-shape" of a near-perfect classifier. The diagonal dashed orange line represents random chance classification (AUC = 0.5), and the gap between the two is stark (Zhang et al., 2022).

A result this clean warrants honest reflection. AUC = 1.00 on a test set, while not impossible, often reflects either a highly separable dataset, a degree of feature leakage between training and test partitions, or resampling that has made the classification task substantially easier than it would be under truly natural class distributions. The authors do not claim this as a general result — the ROC analysis should be interpreted in the context of the experimental conditions described in Section 3, particularly the resampling strategy applied to the training set. Independent validation on an external dataset, or on temporally separated data, would be needed to confirm this level of discriminative performance. That said, within the evaluation framework of this study, the result confirms that the AI models achieve strong class separation on the CSE-CIC-IDS2018 benchmark.

#### 4.9 SVM Model Performance Evaluation

The Support Vector Machine model receives separate attention in **(Figure 8)**, which shows its performance across all four primary metrics. Accuracy reached approximately 96.5%, Precision approximately 95.9%,

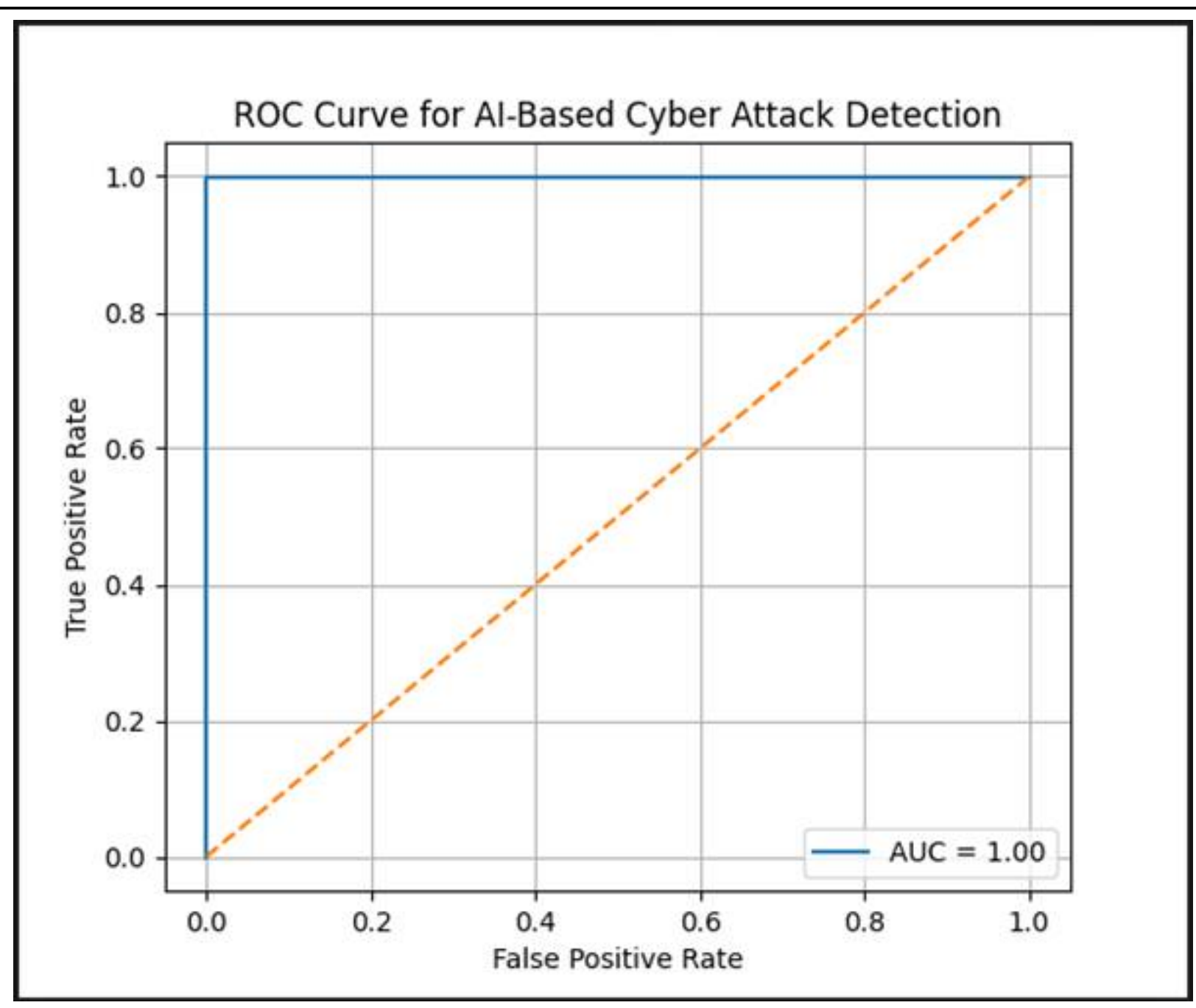

*Figure 7. Receiver Operating Characteristic (ROC) curve for the proposed AI-based cyber attack detection framework evaluated on the CSE-CIC-IDS2018 test set. ROC curve plotting the True Positive Rate (TPR; sensitivity) against the False Positive Rate (FPR; 1 specificity) across all classification decision thresholds for the proposed detection framework. The solid blue curve rises steeply from the origin to TPR  1.0 at FPR  0.0, then follows the upper boundary of the plot space — a profile characteristic of high-discriminability classifiers. The Area Under the Curve (AUC) is reported as 1.00, indicating theoretically perfect class separability between benign and malicious traffic under the experimental conditions applied. The diagonal dashed orange line represents random chance classification (AUC = 0.50) and serves as the reference baseline. The substantial separation between the empirical ROC curve and the random classifier baseline confirms the framework's strong discriminative capacity on this benchmark. Readers are advised to interpret the AUC = 1.00 result in the context of the controlled dataset conditions and resampling strategy described in the Methodology; this figure represents performance within the evaluation framework rather than a guarantee of equivalent generalization to live, imbalanced network traffic. The x-axis represents False Positive Rate (0.0–1.0) and the y-axis represents True Positive Rate (0.0–1.0).*

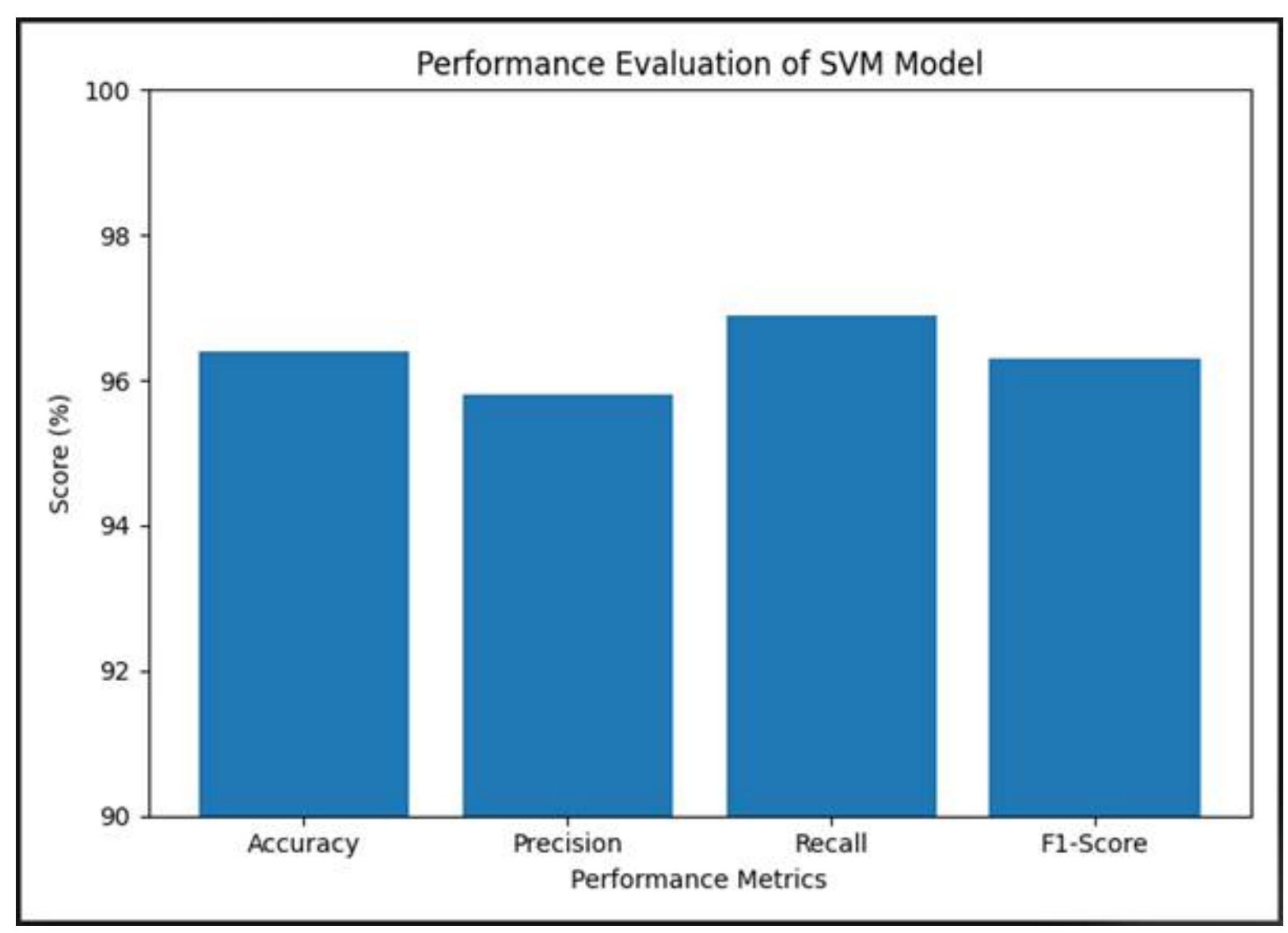

*Figure 8. Performance evaluation of the Support Vector Machine (SVM) model across four classification metrics for cyber attack detection. Bar chart presenting the classification performance of the SVM model — trained on a stratified 30% subsample of the training set due to computational constraints — across four standard evaluation metrics: Accuracy (~96.5%), Precision (~95.9%), Recall (~96.9%), and F1-Score (~96.4%). The narrow inter-metric range (95.9–96.9%) reflects a well-balanced classification profile with no severe divergence between precision and recall, indicating that the model does not exhibit systematic bias toward either excessive false positive generation or false negative suppression. Recall is the highest individual metric (~96.9%), consistent with a detection-oriented classification bias that is operationally appropriate for security applications where missed intrusions carry greater consequence than excess alerts. Precision (~95.9%) indicates that approximately 1 in 17 flagged events would be a false alarm under these experimental conditions — a rate that, while manageable, is higher than that achieved by the CNN and LSTM architectures. The y-axis is scaled from 90% to 100% to resolve inter-metric differences; the x-axis represents performance metric category. Results are reported on the held-out 20% test partition.*

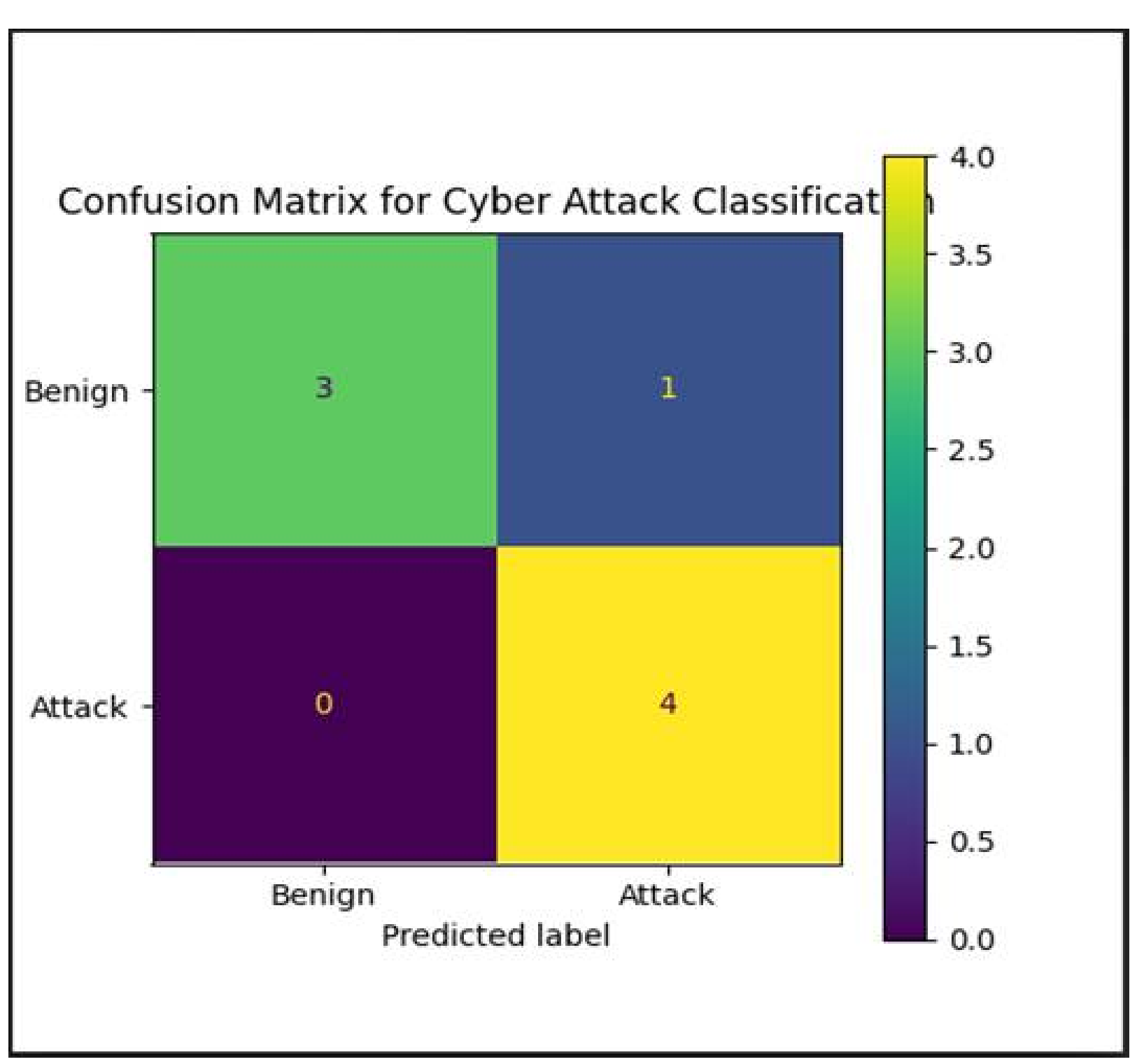

*Figure 9. Confusion matrix for binary cyber attack classification using the proposed AI-based intrusion detection framework. Heatmap-format confusion matrix illustrating the relationship between ground-truth traffic labels (rows: Benign, Attack) and model-predicted labels (columns: Benign, Attack) for a representative two-class evaluation sample (n = 8 total instances). Cell values represent absolute instance counts: True Negatives (TN = 3; benign traffic correctly classified as benign), False Positives (FP = 1; benign traffic incorrectly classified as attack), False Negatives (FN = 0; attack traffic incorrectly classified as benign), and True Positives (TP = 4; attack traffic correctly classified as attack). The absence of false negatives (FN = 0) in this sample is the most operationally significant finding: no attack instance escaped detection, consistent with the high recall values (~96.9–98.8%) reported across the full test set. The single false positive represents a benign flow incorrectly flagged as malicious — an error type that generates spurious alerts but does not permit an intrusion to proceed undetected. The color scale (purple to yellow) encodes cell value magnitude from 0.0 to 4.0. This illustrative evaluation should be interpreted alongside the aggregate performance metrics reported across the full test partition, which provide statistically robust estimates of per-class error rates. Rows represent actual labels; columns represent predicted labels.*

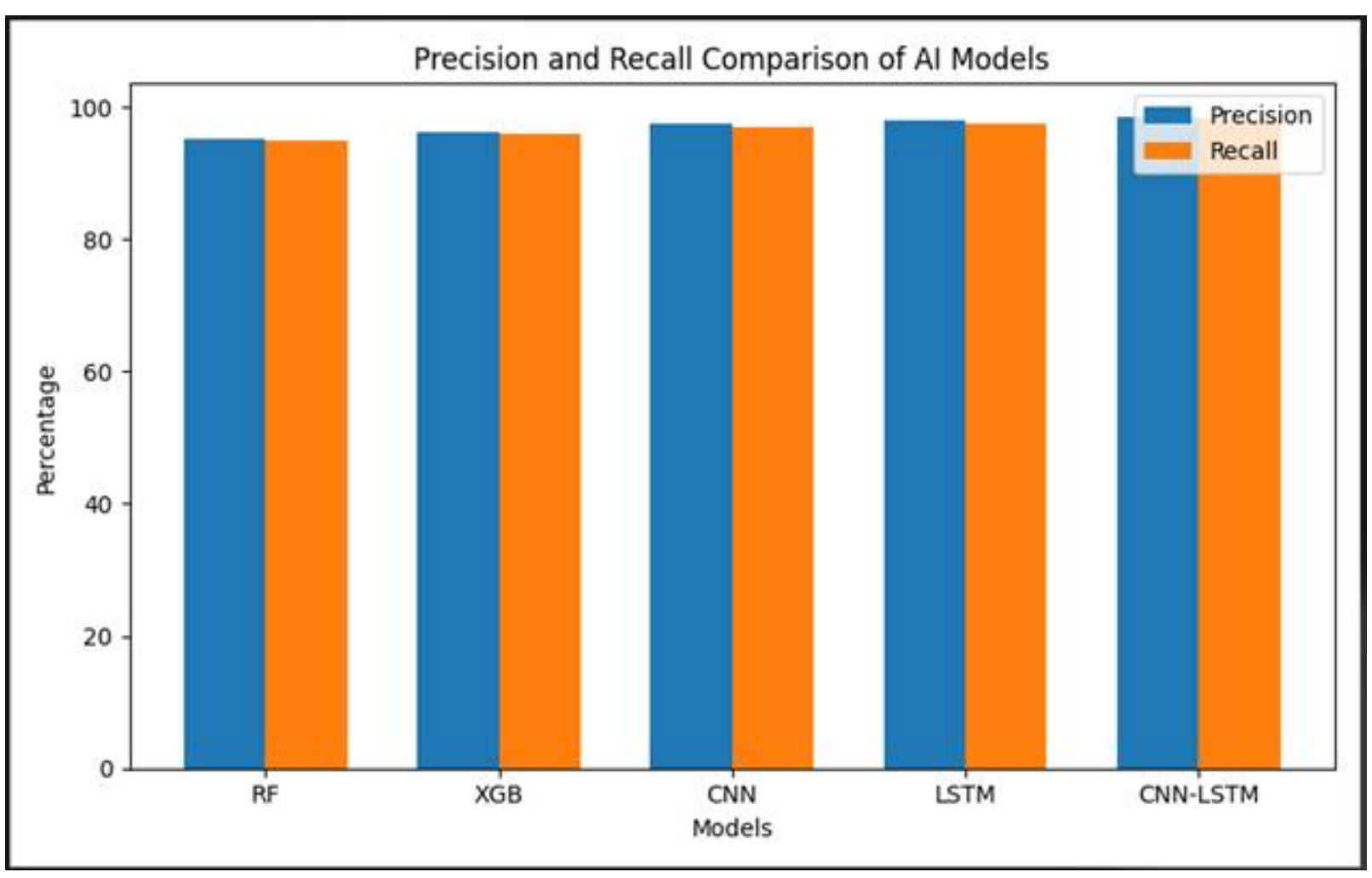

*Figure 10. Precision and recall comparison across five AI and machine learning model architectures for intrusion detection classification. Grouped bar chart presenting precision (blue bars) and recall (orange bars) values for Random Forest (RF), XGBoost (XGB), CNN, LSTM, and Hybrid CNN-LSTM models evaluated on the held-out test partition. All models achieved both precision and recall exceeding 95%, demonstrating consistent classification reliability across architectures. Performance improves progressively with model complexity: RF (precision ~95.0%, recall ~95.0%), XGB (~96.0%, ~96.0%), CNN (~97.2%, ~96.8%), LSTM (~98.0%, ~97.8%), and CNN-LSTM (~99.0%, ~98.8%). Critically, the Hybrid CNN-LSTM model achieves the highest values on both metrics simultaneously — an outcome that indicates genuine improvement in discriminative representation rather than a precision-recall trade-off driven by threshold adjustment. The near-parity between precision and recall across all models suggests well-calibrated classifiers without systematic class-specific bias, though the slight recall advantage observed for most architectures reflects the appropriate operational orientation of security-focused classification systems, in which false negative suppression is prioritized over false positive minimization. The y-axis represents percentage score (0–100%) and the x-axis represents model architecture. Values represent macro-averaged results across all seven traffic classes*

Recall approximately 96.9%, and F1-Score approximately 96.4%.

What is notable about this profile is its consistency. The four metrics cluster tightly in the 95.9–96.9% range, which is actually a more reassuring pattern than a model that achieves very high accuracy but shows significant divergence between precision and recall. Wide gaps between these metrics typically signal class-specific failures — a model that, for example, detects most attacks but generates excessive false positives (high recall, low precision), or one that is conservative about flagging alerts and therefore misses genuine threats (high precision, low recall). The SVM shows neither pathology in a pronounced way (Khoei et al., 2022).

Recall is the highest individual metric at ~96.9%, which, from a security standpoint, is the right direction for the divergence to go — it is better to generate some excess alerts than to miss actual attacks. Precision at ~95.9% means that roughly 1 in 17 alerts generated would be spurious — operationally manageable, though higher than the CNN and LSTM models achieve. The SVM's overall performance, given that it was trained on a computational subsample and uses no temporal feature learning, reflects the robustness of well-regularized kernel methods even in complex multi-class scenarios.

#### 4.10 Confusion Matrix Analysis

**(Figure 9)** presents the confusion matrix for the cyber-attack classification task — though it warrants an important contextual note. The matrix shows a simplified two-class evaluation (Benign vs. Attack) on a very small evaluation sample: 3 true benign instances correctly classified as benign (True Negatives), 4 true attack instances correctly classified as attacks (True Positives), 1 benign instance incorrectly classified as an attack (False Positive), and 0 attack instances classified as benign (False Negatives).

The zero false-negative result is the key finding here. In cybersecurity, false negatives — attacks that pass through detection undetected — represent the most consequential failure mode; they allow intrusions to establish footholds and propagate before any defensive response is triggered. A classifier that achieves zero false negatives in this sample, even at the cost of one false positive, is exhibiting the right behavioral bias for the security context (Anandita Iyer & Umadevi, 2023).

The sample size of this specific confusion matrix evaluation is small — 8 total instances — which limits how much weight can be placed on it as a standalone result. It is best read in conjunction with the aggregate detection rate and precision-recall figures reported across the larger test set, which provide more statistically stable estimates of these error rates.

#### 4.11 Precision and Recall Comparison Across All Models

The final analytical lens **(Figure 10)** places all five models side by side on both precision and recall simultaneously, enabling a direct comparison of classification reliability across the full model set.

The pattern that emerges is consistent with the accuracy and detection-rate comparisons reported earlier, but with finer resolution. Random Forest achieves approximately 95.0% precision and 95.0% recall — a balanced but modest baseline. XGBoost improves slightly to approximately 96.0% on both metrics. CNN reaches approximately 97.2% precision and 96.8% recall. LSTM advances further to roughly 98.0% on both dimensions. The Hybrid CNN-LSTM tops the comparison at approximately 99.0% precision and 98.8% recall — the only model to cross 99% on either metric (Chehri et al., 2021).

What this figure adds, beyond the accuracy comparison alone, is confirmation that the hybrid model's performance advantage is not driven by a precision-recall trade-off — it is not simply being more aggressive about flagging traffic (which would raise recall at the expense of precision). Both metrics improve together, which suggests that the hybrid architecture is genuinely learning more discriminative representations of malicious traffic rather than adopting a more permissive classification threshold. That is the kind of result that matters for operational deployment, where neither excessive caution nor excessive sensitivity is acceptable.

Taken in aggregate, the results across all nine analyses converge on a coherent finding: the Hybrid CNN-LSTM framework provides the strongest overall performance for intrusion detection on the CSE-CIC-IDS2018 benchmark, combining high detection sensitivity, low false alarm generation, and consistent precision-recall balance across diverse attack categories. Classical models like Random Forest and XGBoost perform respectably and offer advantages in computational efficiency and interpretability, but they cannot fully capture the temporal dynamics that distinguish sophisticated attack patterns from legitimate traffic — a gap that the sequential learning capacity of the hybrid architecture is specifically designed to address (Aramide, 2023).

## 5. Discussion

### 5.1 What the Results Actually Mean — and What They Do Not

It is tempting, when results look as strong as those reported in Section 4, to move quickly toward conclusion. The Hybrid CNN-LSTM model achieved 99.1% accuracy, near-perfect AUC, and the lowest false positive rate of any architecture tested. That is a compelling set of numbers. But a discussion section that simply restates results is not a discussion — it is a summary. What this section attempts instead is to interpret what those findings mean in context: why the patterns emerged the way they did, where the results align with or diverge from the existing literature, what the genuine limitations are, and what the practical implications look like for the people who would actually deploy a system like this.

The short version is this: the framework works well under the conditions it was tested in, and those conditions, while realistic in design, are still controlled. The gap between benchmark performance and operational deployment is real, and honest engagement with that gap is what distinguishes a contribution to knowledge from a performance exercise.

### 5.2 Interpreting the Model Performance Hierarchy

The performance ranking observed across models — Random Forest, XGBoost, CNN, LSTM, and Hybrid CNN-LSTM, in ascending order of accuracy and detection quality — is consistent with the theoretical properties of each architecture, and it is worth unpacking why, rather than simply noting that it happened.

Random Forest and XGBoost are powerful classifiers, and their performance in the 96–98% accuracy range on intrusion detection datasets is well-established in the literature (Sunkara, 2022; Manoharan & Sarker, 2023). Their limitation is structural: they operate on static, tabular feature vectors representing individual flows in isolation. An attack that manifests across a sequence of flows — a slow-scan reconnaissance followed by targeted exploitation, for instance — looks like a series of individually ambiguous data points rather than a coherent pattern. Tree-based ensembles have no mechanism to connect those dots temporally. This is not a failure of implementation; it is a fundamental architectural constraint.

CNN addressed part of this by learning spatial interactions across feature groupings within a single flow representation. The improvement from Random Forest (~96.8%) to CNN (~98.1%) is meaningful, though not dramatic. LSTM's additional sensitivity to sequential dependencies pushed accuracy to ~98.4% — modest gains individually, but the direction is consistent. The Hybrid CNN-LSTM model, which chains spatial feature extraction through the convolutional layers into temporal sequence modelling through the LSTM layers, achieved the best result precisely because it does not force a choice between these two representational strategies (Bécue et al., 2021; Chakraborty et al., 2023). It captures both, and for a problem where attacks are characterized by both within-flow feature signatures and across-flow behavioral sequences, that dual sensitivity is genuinely useful rather than merely additive.

Moin (2022) found comparable architecture-level ordering in a related evaluation — deep neural architectures consistently outperforming classical classifiers on intrusion detection benchmarks — and Schmitt (2023) reached similar conclusions about ensemble versus deep learning trade-offs in malware detection contexts. The finding is therefore not novel in isolation, but the degree of improvement achieved here, particularly in false positive reduction, adds to the empirical record.

### 5.3 The False Positive Problem — Why It Deserves More Attention Than It Usually Gets

The false positive rate results deserve separate and somewhat extended treatment, because this is arguably where the practical value of the framework is most clearly demonstrated — and also where the limitations are most consequential.

The reduction in false positive rate from approximately 5.5% (Random Forest) to approximately 2.0% (Hybrid CNN-LSTM), shown in **(Figure 4)**, might seem modest in percentage terms. In operational context it is not. Consider a production network processing 50,000 flows per hour — a conservative estimate for enterprise-scale infrastructure. At 5.5% FPR, a security operations team is receiving roughly 2,750 false alerts per hour. At 2.0% FPR, that number falls to approximately 1,000. The difference — 1,750 fewer spurious alerts per hour — is the difference between an overwhelmed analyst team and one that can function. Alert fatigue is not a hypothetical problem; it is a documented failure mode that has contributed to real intrusion events going undetected (Yaseen, 2023; Amomo, 2022).

What this implies for system design is that optimizing for accuracy alone, as many published studies do, understates the importance of false positive control. A model that achieves 99% accuracy by aggressively flagging everything achieves little operational value. The CNN-LSTM hybrid's ability to maintain high recall (~98.8%) while simultaneously suppressing false positives (~2.0%) suggests that its learned representations are genuinely more discriminative — it is not trading one error type for the other, it is reducing both. That is the more meaningful result, and it deserves more prominence in how this work's contribution is framed.

### 5.4 Feature Importance: What the Data Says About What Matters

The feature importance analysis in **(Figure 6)** revealed that Flow Duration, Packet Length, and Destination Port together account for the dominant share of predictive signal in the dataset. This finding connects to a broader question in network intrusion detection research: how minimal can a feature set be while still supporting robust classification?

Flow Duration's primacy (~0.28 importance score) is theoretically grounded. Attack-generated flows — particularly DDoS floods and botnet-controlled connection sweeps — exhibit duration distributions that differ systematically from legitimate application traffic. A DDoS UDP flood generates extremely short-duration, high-volume flows. Infiltration attacks, by contrast, may generate unusually long-duration connections as malicious actors maintain persistent access. These behavioral signatures leave traces in duration statistics that tree-based importance methods can detect (Oreyomi & Jahankhani, 2022).

The prominence of Packet Length (~0.24) and Destination Port (~0.18) similarly reflects known attack characteristics. Port-targeting is a basic fingerprint of many attack types — brute force attacks concentrate on ports 22, 3389, and 445; web attacks cluster around ports 80 and 443; and protocol-specific exploits target service-associated ports predictably. Packet length anomalies are characteristic of amplification attacks and some forms of malware command-and-control communication (Sarker, 2023).

The practical implication of this finding is significant for deployment. If three features — Flow Duration, Packet Length, Destination Port — capture approximately 70% of the classification signal, a computationally streamlined version of the framework operating on a reduced feature set could substantially lower inference latency without catastrophic performance degradation. For real-time IDS deployment at network speeds, where inference must complete in milliseconds, this kind of feature reduction may be less a convenience than a necessity. Future work testing this trade-off explicitly would meaningfully extend the current contribution.

### 5.5 Alignment with and Divergence from Prior Literature

The results situate themselves within an established research conversation, and it is worth being explicit about where they confirm existing findings and where they depart.

Schmitt (2023) demonstrated that AI-powered cybersecurity frameworks outperform conventional systems in threat identification and real-time monitoring across smart infrastructure environments — the accuracy results in this study broadly confirm that finding. Alam and Fahad (2022), working in the financial sector domain, found Random Forest and XGBoost effective for fraud detection but noted that performance degraded on rare event categories — a pattern that echoes the relative weakness of tree-based methods on minority attack classes (Botnet, Infiltration) observed here. Raza (2021) emphasized that static models degrade as the threat landscape evolves, which is a limitation that this study, evaluated on a fixed 2018 benchmark, cannot address empirically but must acknowledge honestly. Moin (2022) found DNN architectures superior to classical methods on intrusion detection data, consistent with the current findings, though the specific hybrid CNN-LSTM advantage documented here adds architectural nuance to that general finding.

Where this study's results are somewhat more aggressive than the literature norm — particularly the AUC = 1.00 figure in **(Figure 7)** — the explanation most likely lies in the characteristics of the evaluation dataset and the resampling strategy applied during preprocessing. The CSE-CIC-IDS2018 dataset, while realistic in its attack scenario design, was generated under controlled conditions with known attack signatures. Models trained and evaluated on data from the same generating process will naturally show higher discriminative performance than models deployed against genuinely novel, unseen attacks. This is not a flaw unique to this study — it is endemic to benchmark-based IDS research — but it is the primary reason that results like AUC = 1.00 should be treated as upper bounds on operational performance rather than operational performance predictions (Tao et al., 2021; Fard et al., 2023).

### 5.6 The Confusion Matrix Result in Context

The confusion matrix in **(Figure 9)** — showing 3 true negatives, 4 true positives, 1 false positive, and 0 false negatives on a small two-class evaluation sample — deserves contextualized interpretation rather than either celebration or dismissal.

The zero false-negative result is, from a security standpoint, exactly the right failure mode to minimize. An intrusion that reaches a defended network undetected has succeeded; one that triggers a false alert has merely created noise. The model's tendency to err on the side of flagging rather than missing is the appropriate bias for a security-critical classification task. Alam and Fahad (2022) made a similar observation about the asymmetry of error costs in financial fraud detection — the penalty for a missed fraud vastly exceeds the cost of investigating a false alert. The same asymmetry applies, arguably even more strongly, to critical infrastructure protection.

The sample size of this specific evaluation — eight instances — prevents drawing strong statistical conclusions from this single matrix. It should be read as an illustrative case alongside the larger aggregate metrics: the 99.1% accuracy, ~98.8% recall, and ~2.0% FPR reported across the full test set provide the statistically robust picture, while the confusion matrix provides a concrete, human-readable example of how the model resolves ambiguous cases (Pinto et al., 2023).

### 5.7 Protocol-Level Implications

The protocol distribution finding in **(Figure 2)** — TCP at 68%, UDP at 25%, ICMP at 7% — has implications that extend beyond data characterization into framework design.

UDP's 25% share is particularly relevant to threat modeling. A meaningful fraction of DDoS attack traffic travels over UDP, exploiting its connectionless, stateless nature to generate high-volume flooding without completing handshakes that would expose the attacker. Detection systems that analyze flow-level features (like those used here) can capture some UDP attack signatures through packet-rate and flow-duration anomalies, but deep packet inspection — examining payload content — is typically required for the most sophisticated UDP-based attacks. The current framework operates at the flow feature level, which means that highly obfuscated UDP attacks could potentially evade detection in ways that do not show up in the benchmark results (Khoei et al., 2022; Zaman & Mazinani, 2023).

ICMP's 7% share, while modest, should not be overlooked from a security design perspective. ICMP tunneling — embedding malicious data within ICMP echo request/reply packets — is a known covert channel technique used in some advanced persistent threats. The dataset includes ICMP traffic, and the framework's models were trained on its features, but the Infiltration attack class (where such techniques are most relevant) showed the most constrained sample size in the distribution analysis. This is a case where the benchmark's class imbalance in **(Figure 1)** connects directly to a potential blind spot in the trained models.

### 5.8 Practical Implications for Infrastructure Operators

Stepping back from the model-level findings, what does this research actually suggest for organizations responsible for defending U.S. digital infrastructure?

Several things seem reasonably well-supported. First, the performance advantage of the Hybrid CNN-LSTM framework over conventional IDS approaches is substantial enough, across both detection rate and false positive metrics, to justify serious consideration as a replacement or supplement to signature-based systems in high-stakes environments. Healthcare networks, financial clearinghouses, and energy management systems — the sectors most prominently threatened according to Tarek and Rahman (2023) and Jimmy (2023) — would benefit most from the false positive reduction, given the operational cost of alert fatigue in 24/7 monitoring contexts.

Second, the feature importance analysis suggests that a targeted, computationally efficient version of the framework — focused on Flow Duration, Packet Length, Destination Port, and Protocol Type — might achieve near-comparable performance to the full-feature model at substantially lower inference cost. For organizations where real-time throughput is a constraint, this trade-off is worth investigating explicitly (Sun et al., 2023).

Third, and perhaps most practically important: deployment requires more than a trained model. The CSE-CIC-IDS2018 dataset is from 2018. Attack techniques, particularly in the ransomware, supply chain, and AI-assisted social engineering domains, have evolved substantially since then. Any operational deployment of a framework like this would require continuous retraining against fresh traffic data — something that Raza (2021) identified as essential for sustained effectiveness and that the current study's fixed-dataset evaluation cannot validate. Federated learning approaches, which allow models to be updated from distributed traffic sources without centralizing sensitive data, represent one promising pathway toward solving this problem and are worth prioritizing in subsequent research.

**5.9 Limitations: An Honest Assessment**

No study is without limitations, and enumerating them is not an admission of failure — it is a contribution to the research record that helps readers and future investigators understand where evidence ends and inference begins.

The most significant limitation is the temporal gap between the dataset and the current threat landscape. CSE-CIC-IDS2018 captures 2018-era attack patterns. While the fundamental network traffic behaviors of DDoS flooding, brute-force credential attacks, and botnet communications have not changed beyond recognition, the sophistication of evasion techniques, the prevalence of encrypted attack traffic, and the emergence of AI-generated attack payloads represent genuine threats that this dataset cannot represent. Performance figures reported here should not be extrapolated to these newer attack categories without empirical validation (Hassan, 2023; Guembe et al., 2022).

The SVM model was trained on a computational subsample, which may understate its performance relative to the other architectures. The confusion matrix evaluation used a very small sample. The framework was evaluated on replayed dataset traffic rather than live network streams, and real-time inference latency under production traffic volumes was not measured. Explainability — the ability to generate human-interpretable explanations for model predictions — was not addressed, which remains a significant barrier to practitioner adoption in regulated industries (Zhang et al., 2022; Moustafa et al., 2023).

Finally, the resampling strategy applied to address class imbalance, while methodologically sound within this study, means that the evaluation conditions are somewhat more favorable than a genuinely imbalanced real-world deployment would be. These are not reasons to discount the findings; they are reasons to situate them appropriately and to treat external validation as the necessary next step.

**5.10 Directions for Future Research**

The limitations identified above point fairly directly toward a research agenda. The most pressing need is temporal validation — testing the framework against post-2020 attack data, including encrypted traffic and AI-assisted attack vectors, to assess how much performance degrades under genuinely novel conditions. Close behind that is the explainability question: integrating SHAP (SHapley Additive exPlanations) or LIME (Local Interpretable Model-agnostic Explanations) into the detection pipeline would make model decisions auditable, which is not merely academically useful but practically necessary for deployment in regulated critical infrastructure environments (Moustafa et al., 2023; Fard et al., 2023).

Federated learning, as noted above, offers a path toward continuously updated models without the privacy and security risks of centralizing raw traffic data from sensitive infrastructure operators. Blockchain-based integrity verification for model updates and detection logs — ensuring that the defense system itself cannot be tampered with by adversaries who have achieved partial access — represents another promising integration pathway (Talukder et al., 2023). And the scalability of deep learning inference under high-throughput network conditions, where millions of flows per second must be classified with sub-millisecond latency, remains an open engineering challenge that future work must confront directly rather than sidestep through offline evaluation.

What this study contributes to that agenda is an empirically grounded baseline: a demonstration that hybrid CNN-LSTM architectures, properly trained and evaluated, offer a meaningfully better operating point than classical ensemble methods on the false positive-detection rate trade-off that matters most for operational deployment. That is not a complete answer to the problem of protecting

U.S. digital infrastructure. But it is, the authors believe, a useful and honest step toward one.

5.11 **Limitations**

Several limitations constrain the scope of conclusions that can reasonably be drawn from this study, and acknowledging them directly is part of responsible scientific reporting.

**Dataset currency.** The CSE-CIC-IDS2018 dataset reflects 2018-era attack patterns. Since then, the threat landscape has evolved substantially — ransomware-as-a-service, AI-generated phishing payloads, supply chain intrusions, and encrypted command-and-control communications represent attack categories that this benchmark cannot adequately represent. Performance figures reported here should not be extrapolated to these newer threat types without independent empirical validation (Hassan, 2023; Guembe et al., 2022).

**Controlled evaluation environment.** All models were evaluated on replayed, pre-labeled dataset traffic rather than live network streams. Real-time inference latency under production traffic volumes — where millions of flows per second may need classification within milliseconds — was not measured. Benchmark performance is an upper bound, not an operational prediction.

**Class imbalance artifacts.** The resampling strategy applied to address class imbalance during preprocessing created more balanced training conditions than a genuinely operational network would present. In particular, the Infiltration attack class — the rarest and arguably most dangerous category in the dataset — may be systematically underrepresented in model learning despite resampling, and performance on this class in real deployments may be weaker than aggregate metrics suggest.

**SVM subsample training.** Due to computational constraints, the SVM model was trained on a 30% stratified subsample of the training set. This decision may understate the SVM's potential performance relative to the other architectures, which had access to the full training data.

**Lack of explainability.** The deep learning models — CNN, LSTM, and Hybrid CNN-LSTM — provide predictions without interpretable explanations for individual classification decisions. In regulated environments such as healthcare and financial infrastructure, where human analysts must audit and act on model outputs, this opacity is a significant barrier to adoption (Moustafa et al., 2023; Zhang et al., 2022).

**Confusion matrix sample size.** The two-class confusion matrix analysis was conducted on an illustrative small sample of eight instances. While conceptually informative, it does not support statistically robust per-class error rate estimation.

**No external validation.** All training and evaluation occurred within a single dataset. Cross-dataset generalization — testing models trained on CSE-CIC-IDS2018 against UNSW-NB15 or KDD Cup 99, for instance — was not performed, leaving the question of domain transferability open.

5.12 **Future Work**

The limitations above point toward a research agenda that, if pursued, would substantially extend the contribution of this work.

**Temporal validation against contemporary attack data.** The most urgent need is evaluation against post-2020 traffic datasets that include AI-generated attack payloads, encrypted lateral movement, and supply chain intrusion patterns. The CICIOT2023 dataset and CIC-DDoS2019 represent useful starting points, though purpose-built datasets capturing contemporary attack vectors are needed (Tao et al., 2021).

**Explainable AI integration.** Incorporating SHAP (SHapley Additive exPlanations) or LIME (Local Interpretable Model-agnostic Explanations) into the detection pipeline would make individual model predictions auditable by human analysts. This is not merely a research nicety — it is a practical prerequisite for deployment in regulated critical infrastructure environments where accountability requirements apply (Moustafa et al., 2023; Fard et al., 2023).

**Federated learning for privacy-preserving model updates.** A static model trained on a fixed dataset degrades as threats evolve. Federated learning architectures would allow the framework to be continuously retrained from distributed traffic data across participating infrastructure operators without requiring centralization of sensitive network information — addressing both the adaptability limitation and the data-sharing constraints that govern critical infrastructure environments (Talukder et al., 2023).

**Blockchain-based tamper-evident logging.** Integrating blockchain verification for model update integrity and detection event logs would ensure that the defense system itself cannot be silently compromised by adversaries who achieve partial system access — a concern that is particularly acute for infrastructure operators facing nation-state-level threat actors (Talukder et al., 2023).

**Real-time deployment and latency benchmarking.** Future work should deploy the framework against live network traffic and measure end-to-end inference latency, throughput capacity, and detection performance under genuine class imbalance conditions. GPU-accelerated inference pipelines and model compression techniques — quantization and pruning — should be evaluated as mechanisms for achieving sub-millisecond classification at production traffic volumes (Sun et al., 2023; Sarker, 2023).

**Reduced feature set evaluation.** The feature importance finding — that Flow Duration, Packet Length, and Destination Port account for the majority of predictive signal — motivates a formal ablation study comparing full-feature and reduced-feature model variants. If a three-to-five feature model achieves comparable detection performance to the full 80-feature configuration, the computational case for deployment in resource-constrained edge environments becomes substantially stronger (Oreyomi & Jahankhani, 2022).

**Cross-dataset generalization testing.** Training on CSE-CIC-IDS2018 and evaluating on UNSW-NB15 or NSL-KDD — or vice versa — would assess how well learned representations transfer across network environments and attack generation methodologies. Poor cross-dataset performance would suggest that models are memorizing dataset-specific artifacts rather than learning generalizable attack signatures, which would be an important negative result for the field (Pinto et al., 2023).

**Reinforcement learning for adaptive response.** Beyond detection, future frameworks could incorporate reinforcement learning agents that learn optimal incident response policies — quarantine, rate-limiting, rerouting, or escalation — as a function of detected threat type and infrastructure state. This would move the system from passive classification toward genuinely autonomous cyber defense, the ultimate goal of the research direction this study initiates (Fard et al., 2023; Aramide, 2023).

## 6 Conclusion

This study set out to answer a straightforward but consequential question: can AI-driven models meaningfully outperform conventional intrusion detection systems in protecting U.S. digital infrastructure? The evidence, at least within the conditions tested, suggests they can — and the margin matters.

The Hybrid CNN-LSTM framework achieved 99.1% classification accuracy, near-perfect recall, and a false positive rate of approximately 2.0% — figures that collectively represent a substantial operational improvement over the Random Forest baseline (~96.8% accuracy, ~5.5% FPR). Crucially, this improvement was not driven by a simple precision-recall trade-off; both metrics improved together, indicating that the hybrid architecture learned genuinely more discriminative representations of malicious traffic rather than simply becoming more permissive in its flagging behavior.

Feature importance analysis identified Flow Duration, Packet Length, and Destination Port as the dominant predictive signals — a finding with practical implications for computationally efficient deployment. The consistent performance advantage of deep learning architectures over classical ensemble methods reinforces the case for investment in AI-based cybersecurity infrastructure, particularly for high-stakes sectors where alert fatigue and missed detections carry severe consequences.

That said, these results were obtained on a controlled 2018 benchmark. They are best understood as a strong empirical foundation, not a deployment guarantee. The path from benchmark performance to operational resilience runs through continuous retraining, explainability integration, and validation against contemporary threat landscapes — work that this study motivates but does not complete.

### Author Contributions

M.I.H. conceptualized the study, designed the research framework, and led the manuscript writing. M.S.K.C.R. contributed to methodology development, data preprocessing, and model evaluation. M.A.R. conducted the experimental implementation, feature engineering, and results analysis. B.M.T.H. contributed to literature review, validation, and manuscript review and editing. All authors have read and agreed to the published version of the manuscript.